\documentclass[
pof,
aip,
reprint,
amsmath,amssymb,
sd,
]{revtex4-1}
\usepackage[T1]{fontenc}
\usepackage[utf8]{inputenc}
\usepackage{amsmath}
\usepackage{amssymb}
\usepackage{enumitem}
\usepackage{graphicx}
\usepackage{refstyle}
\usepackage{multirow}

\usepackage[
citecolor=blue,
colorlinks,
linkcolor=blue,
urlcolor=blue,
]{hyperref}

\usepackage{xcolor}
\usepackage{mathtools}
\DeclarePairedDelimiter{\ceil}{\lceil}{\rceil}
\usepackage{amsfonts}	
\usepackage{float}		
\usepackage{longtable}
\usepackage{array}
\usepackage{hyperref}

\newcommand{\mb}[1]{\mathbf{#1}}

\newcommand{\vex}{\varepsilon(x)}

\begin{document}
	\preprint{AIP/123-QED}
	\title{Bridging Inertial and Dissipation Range Statistics in Rotating Turbulence}
	\author{Shailendra K. Rathor}
	\email{skrathor@iitk.ac.in}
	\affiliation{Department of Physics, Indian Institute of Technology Kanpur, Uttar Pradesh 208016, India}
	\author{Manohar Kumar Sharma}
	\email{kmanohar@iitk.ac.in}
	\affiliation{Department of Physics, Indian Institute of Technology Kanpur, Uttar Pradesh 208016, India}
	\author{Samriddhi Sankar Ray}
	\email{samriddhisankarray@gmail.com}
	\affiliation{International Centre for Theoretical Sciences, Tata Institute of Fundamental Research, Bangalore 560089, India}
	\author{Sagar Chakraborty}
	\email{sagarc@iitk.ac.in}
	\affiliation{Department of Physics, Indian Institute of Technology Kanpur, Uttar Pradesh 208016, India}
	\begin{abstract}
		
		We investigate the connection between the
		inertial range and the dissipation range statistics of rotating
		turbulence through detailed simulations of a helical shell
		model and a multifractal analysis. In particular, by using the
		latter, we find an explicit relation between the (anomalous)
		scaling exponents of equal-time structure functions in the
		inertial range in terms of the generalised dimensions
		associated with the energy dissipation rate. This theoretical
		prediction is validated by detailed simulations of a helical
		shell model for various strengths of rotation from where the
		statistics of dissipation rate, and thus the generalised
		dimensions, as well as the inertial range, in particular the
		anomalous scaling exponents, are extracted.  Our work also
		underlines a surprisingly good agreement---such as in the
		spatial structure of the energy dissipation rates and the
		decrease in inertial range intermittency with increasing
		strengths of rotation---between solutions of the Navier--Stokes
		equation in a rotating frame with those obtained from
		low-dimensional, dynamical systems such as the shell model
		which are not explicitly anisotropic. Finally, we perform
		direct numerical simulations of the Navier--Stokes equation,
		with the Coriolis force incorporated, to confirm the robustness of the
		conclusions drawn from our multifractal and shell model
		studies.

	\end{abstract} 
	\keywords{turbulence; rotating; multifractal; shell models} 
	\maketitle
	
	\section{Introduction}
	
	Turbulent flows are amongst the more well-known problems where the use
	of standard tools of statistical physics and analysis has met with
	limited success. One of the factors contributing to this is the
	intermittent nature of the flow~\cite{Frisch1995,Lohse1993} which, in
	turn, leads to non-Gaussian distributions of observables such as
	velocity gradients as well as the multiscaling of (suitably-defined)
	$q$-th order moments of the spatial increments of the velocity field:
	Higher-order moments (and their exponents $\zeta_q$) are not trivially
	(linearly) related to lower-order
	moments~\cite{Kolmogorov1941a,Kolmogorov1941b,Kolmogorov1962}.
	{We recall that these moments, referred to as
	equal-time structure functions, and the associated exponents, for a velocity field ${\bf v}({\bf x})$ are
	defined as $S_q(r) \equiv \langle \delta v \,^q \rangle \sim
	r^{\zeta_q}$, where $\delta v \equiv [{\bf v}({\bf x} + {\bf r}) - {\bf
	v}({\bf x})]\cdot{\bf \hat r}$. The angular brackets denote an average
	over the spatial points ${\bf x}$; the separation ${\bf r}$ over which
	this velocity difference is taken (and then projected along that vector
	${\bf \hat r}$ to obtain the longitudinal component) is assumed to be
	within the inertial range of scales of turbulence.  In other words $L
	\gg r \gg \eta$, where $L$ is the largest scales where energy is
	injected and $\eta$ is the smallest (Kolmogorov) length scales where
	energy is dissipated.} Several experiments and direct numerical
	simulations (DNSs) of the three-dimensional, incompressible
	Navier-Stokes equation for fully-developed, statistically homogeneous
	and isotropic turbulence have now established beyond doubt that not
	only are distributions of velocity-gradients and fluid acceleration
	characterised by non-Gaussianity and fat
	tails~\cite{Kailasnath-1992,Bodenschatz-Nature} but the scaling
	exponents $\zeta_q$ are non-linear, concave, monotonically increasing
	functions of $q$. 
	
	While the inertial range exponents display multiscaling, there is also
	overwhelming experimental and numerical evidence~\cite{Batchelor1949} that the
	energy dissipation rates show strong temporal and spatial
	fluctuations~\cite{Siggia1981} characterized by periods of intense bursts and
	calmness. An important question is to find rigorous estimates for
	the statistics of the energy dissipation rate $\epsilon$ which is, within the
	Kolmogorov picture, on average equal to the constant energy flux across the
	inertial range of scales. A popular candidate to model the statistics of
	$\epsilon$ is to use a  log-normal form to fit the probability distribution
	function of the energy dissipation rate~\cite{Kolmogorov1962}. Subsequently,
	this issue of the statistics of the dissipation rate has been revisited and the
	problems associated with the log-normal
	assumption~\cite{Mandelbrot1972,Kraichnan1974} eventually gave way to fractal
	models~\cite{Mandelbrot1974,Frisch1978} culminating in the Frisch--Parisi
	multifractal formalism~\cite{Ghil1985}. One of the great successes of the
	latter was the rationalization of the observed multiscaling of velocity
	structure functions~\cite{Frisch1995,Pandit-Review}. 
	
	This connection between the statistics of the dissipation and inertial ranges
	has been studied extensively for turbulence which is homogeneous and isotropic.
	However, a similar analysis and its consequences when isotropy is explicitly
	broken is far from obvious. In this work, we explore this question in some
	detail by studying turbulent flows in a setting where isotropy is explicitly
	broken. A natural choice for this is the problem of rotating turbulence~\cite{Greenspan1968,
		Moffat_1983, davidson_2013} whose
	ubiquity ensures that our study is not merely an academic exercise but an
	important addition in areas of fluid dynamics, geophysics, and astrophysics~\cite{Barnes_2001,James_2008,Reun_2017,Aurnou_2015}.
	Indeed there are several examples of turbulent flows which are inevitably
	associated with the Coriolis force that, while doing no work {or additional 
	energy injection, reorganizes the structure of the flow, introduces anisotropy and hence} leads to physics
	quite distinct (often mediated by large-scale columnar vortices) from
	non-rotating turbulent flows~\cite{Smith_1999, Muller_2007, Mininni2009a,Bartello_1995,metais_1996, Yarom_2013}. Thus, unsurprisingly, the last few decades have
	seen major theoretical and experimental studies on the different Lagrangian and
	Eulerian aspects of rotating turbulence~\cite{bartello_1994,Bartello_1995,metais_1996,
		godeferd_1999,Zeman1994, Hattori2004, Muller_2007, Mininni2009a,
		sreenivasan_2008, Pouquet_2010,Castello_2011,Biferale2016,Maity,ibbetson_tritton_1975,
		hopfinger_1982, Bewley_2007, Morize_2005, moisy_2011, Yarom_2013, Yarom_2017, Sharma2018a, Sharma2018, Sharma2019}.  Surprisingly, though, the issue of
	the statistics of the Eulerian dissipation field (beyond a recent work on
	Lagrangian irreversibility) and its connections with the observed modifications
	of the statistics of the inertial range, quantified via the scaling exponents
	of the equal-time structure functions, has not been studied in-depth for such
	flows~\cite{Seiwert2008,Thiele2009,Mininni2010a,Imazio2017}. 
	
	In this paper, we tackle this question. In particular,  we (a) show how with increasing rotation rates not
	only does the energy dissipation field appear less intermittent with associated
	changes in its multifractal spectrum{, and} (b) establish a relation between the
	anomalous exponents of equal-time velocity structure functions, measured in the
	inertial range to the R\'enyi scaling exponent obtained from partition
	functions of the dissipation field $\vex$.
	
	Experiments and DNSs are of course primary sources on which theoretical and
	phenomenological ideas are built. Nevertheless, \textit{synthetic} models still
	serve as useful tools to develop insights on the origins of intermittency and
	the curious nature of energy dissipation~\cite{Kraichnan1958,Kraichnan1959,Orszag,Lesieur,Frisch2001,Saha2008,Chakraborty2009,Ray-Review,Ray-PRF2018}. A particularly useful example of this
	is the class of  cascade models known as shell models. Remarkably such models
	which have very little in common (beyond the formal structure) to the
	Navier--Stokes equations, are shown, for homogeneous and isotropic turbulence,
	to mimic the multiscaling of two-point correlation functions with remarkable
	accuracy and hence have, over the years, proved a remarkable testing ground for
	theories of various correlation functions which were inaccessible to full-scale
	simulations or experiments. 

	Rotating turbulent flows are however inherently anisotropic. Therefore, for such 
	systems, understandably, the use of shell models have been sparse. Therefore, in this work, beyond our theoretical 
	(multifractal) analysis, we resort to extensive simulations of a shell model which incorporates the effect 
	of the Coriolis force without explicitly resolving the anisotropy which manifests itself through structural differences
	in the flow in planes parallel and perpendicular to the axis of rotation. Nevertheless, as we find, 
	such a shell model is still robust enough to (a) pick out the relevant Zeman scales and hence the two power-law regimes in the energy spectrum, 
	(b) predict the scaling exponents of the equal-time structure functions (without, of course,  distinguishing the 
	planes parallel and perpendicular to the axis of rotation) in the inertial range with significant accuracy, and (c) allow 
	an extraction of the spatial profile of the energy dissipation rate, through a method due to Lepreti,
	Carbone, and Veltri~\cite{Lepreti2006}. The last of these allows us to extract the generalised dimension $D_q$ for the dissipation 
	rate statistics which, coupled with measurements of $\zeta_q$, serves to independently verify the validity of our theoretical 
	prediction bridging the statistics of the inertial and the dissipation ranges. In the end, we finally resort to full-scale 
	direct numerical simulations of the Navier--Stokes equation, with rotation, as well as results from DNSs reported by others in the past, 
	to validate of the central results of our shell model study. This agreement between the two approaches not only underlines one of the objectives 
	of this work, \textit{viz.}, the extent and usefulness of modeling rotating turbulence as a low-dimensional dynamical systems model, 
	but it also serves as an additional confirmation of the theoretical predictions borne out of the multifractal approach.
	
	\section{A Helical Shell Model for Rotating Turbulence}
	\label{theory}
	We begin with the incompressible ($\mb{\nabla}\cdot \mb{u} = 0$) Navier--Stokes equation for the velocity field $\mb{u}$ of a three-dimensional flow, 
	with density $\rho$ and a kinematic viscosity $\nu$ small enough to generate turbulence, under a solid body rotation $\mb{\Omega}$:
	\begin{equation}
		\frac{\partial \mb{u}}{\partial t} + (\mb{u}.\mb{\nabla}) \mb{u}  =   -\frac{1}{\rho} \mb{\nabla} p + \nu \nabla^2 \mb{u} - 2 \mb{\Omega} \times \mb{u} + \mb{f}.
		\label{eqn:NSE}
	\end{equation}
	The pressure $p$ includes the effect of the centrifugal force and the Coriolis force $-2 \mb{\Omega} \times \mb{u}$ results from the solid body rotation. Furthermore, an external, large-scale, force $\mb{f}$ ensures that the
	(turbulent) flow remains in a statistically stationary state through the injection of an energy $\epsilon = \langle \mb{u}\cdot \mb{f} \rangle$. Unlike non-rotating three-dimensional turbulent flows, helicity plays an important
	role in rotating turbulence; a natural source of helicity is the Coriolis force which results in a helicty injection $\langle 2 \mb{u} \cdot \nabla (\mb{\Omega} \cdot \mb{u}) \rangle$; similarly the external drive can also inject helicity at large-scales via helicity $\langle \omega \cdot \mb{f} + \mb{u} \cdot (\nabla \times \mb{f}) \rangle$. (The angular brackets in these definitions imply suitable averaging in space or in time for non-equilibrium stationary states.)
	
	The solution of Eq.~(\ref{eqn:NSE}) is characterized not only by its Reynolds number $Re$ (as would be the case for non-rotating flows) but by a second non-dimensional number, the Rossby number $Ro = \mb{u}_{\rm
		rms}/(2L\mb{\Omega})$, which is a measure of the relative importance of the Coriolis and inertial terms in flow; $L$ is the characteristic length of the domain ({typically, 2$\pi$ in numerical simulations) and $\mb{u}_{\rm rms}$ is the root-mean-square velocity. }
	
	Rotating turbulent flows in nature or in laboratories 
		have often a non-zero helicity. This non-zero helicity, in turn, affects the statistical properties of the flow principally 
		through a modification of the fluxes. Given that in this paper we bridge inertial to dissipative range statistics, it 
		is useful to avoid possible (non-essential) contributions coming in via a mean helicity. Thus, we 
		are careful in ensuring a zero mean-helicity injection in the model we simulate (see below);
		we have nevertheless checked that small to moderate mean helicities do not change the central results of this paper.

	Shell models are essentially low-dimensional dynamical systems which mimic the
	spectral Navier--Stokes equation without being actually derived from
	it~\cite{Frisch1995,Bohrbook,Biferale-review,Pandit-Review}.  The dynamical
	system is constructed by replacing the three-dimensional Fourier space with a
	one-dimensional logarithmically-spaced shell-space. We associate with each
	shell $n$ in this lattice, corresponding to a wavenumber $k_n = k_0 \lambda^n$,
	a dynamical, complex variable $u_n$ which mimics the velocity increments over a
	scale $r \sim 1/k_n$ in the Navier--Stokes equation. The actual structure of
	this shell-space is determined by the constant $k_0$ and $\lambda$. It is this
	logarithmic construction on a one-dimensional lattice and restricting the
	non-linear interactions to just the nearest and next-nearest neighbours which
	allows shell models to achieve extremely high Reynolds numbers---and hence
	inertial ranges---well beyond those possible through DNSs.  Curiously, shell
	models  seem to give very reliable measurements of the anomalous, due to
	intermittency, scaling exponents of structure functions and the energy
	spectrum; however not much is known about the dissipation statistics of such
	models.  Thus such models have been used extensively in the past for problems
	which ranged from studies of static and dynamic multiscaling in fluid,
	passive-scalar, binary fluids and magnetohydrodynamic
	turbulence~\cite{Biskamp1994,Wirth1996,Basu1998,Frick1998,Mitra2004,Mitra2005,Ray2008,Ray2011,Banerjee2013}, turbulent flows with
	polymer-additives and elastic turbulence~\cite{Benzi2003,Kalelkar2005,Ray2016} as well the equilibrium
	solutions of such dynamical systems~\cite{Ditlevsen1996,Gilbert2002,Tom2017}.
	However, the application of such low-dimensional models for rotating turbulence
	is both sparse and fairly recent~\cite{Chakraborty2010,Hattori2004}.

	To this end, we simulate a helical shell model (to ensure a zero mean helicity for the reasons 
		mentioned before), for rotating turbulence, 
		constructed by decomposing the velocity field in the basis corresponding to the eigenvectors of the curl operator~\cite{Benzi1996,Waleffe1992}:
	\begin{eqnarray}
		\mathbf{u}(\mathbf{x}) & = & \sum_{\mathbf{k}}\mathbf{u}(\mathbf{k}) \exp(i\mathbf{k}\cdot\mathbf{x}) \nonumber \\
		& = & \sum_{\mathbf{k}} [u^+(\mathbf{k}) \mathbf{h}_+ + u^-(\mathbf{k}) \mathbf{h}_- ]\exp(i\mathbf{k}\cdot\mathbf{x}).
	\end{eqnarray}
	Here $\mathbf{u}^\pm$ are the velocity components along the unit eigenvectors $\mathbf{h}_\pm$ of the curl operator 
	$i\mathbf{k} \times \mathbf{h}_\pm = \pm k \mathbf{h}_\pm$.
	Such a decomposition is adapted to a shell model framework to yield the following set of ordinary differential equations 
	\begin{eqnarray}\label{eq1}
		\frac{d}{dt}u_n^\pm = \mathcal{N}^\pm -\nu k_n^2 u_n^\pm + \mathcal{F}_m^\pm -i\Omega u_n^\pm.
	\end{eqnarray}
	The non-linear terms $\mathcal{N}^\pm$ are defined as
	\begin{eqnarray}
		\mathcal{N}^\pm &=& i \left[a k_{n+1} u_{n+2}^{\mp} u_{n+1}^{\pm} + b k_{n} u_{n+1}^{\mp}u_{n-1}^{\pm} \right. \nonumber \\ & &\left. + c k_{n-1} u_{n-1}^{\mp} u_{n-2}^{\mp}\right]^{*}
	\end{eqnarray}
	with real coefficients $a,b$ and $c$, the superscript * denoting a complex conjugate, and 
	the effective velocity associated with each shell  
	$u_n = \sqrt{|u_n^+|^2 + |u_n^-|^2}$. We set the coefficient $a$ to unity and 
	the coefficients $b = -\frac{\lambda^{-1} + \lambda}{\lambda +1}$ 
	and $c = \frac{\lambda^{-1} + 1}{\lambda +1}$ to ensure the conservation ($\nu = 0$) of energy ($a+b+c = 0$) and helicity ($a+b\lambda-c\lambda^2 = 0$).
	The structure of the shell model allows an easy identification of 
	the viscous dissipative term and a forcing term on the 
	$m^{th}$ shell with $\mathcal{F}_m^\pm = \epsilon^\pm (1+i) /(u^\pm_m)^* $; $\epsilon^\pm$ is the energy input rate to the mode $u_m^\pm$ and we choose $m = 3$.
	The helicity injection rate is given by $ k_m (\epsilon^+ - \epsilon^-)$; by choosing $\epsilon^+ = \epsilon^-$ we ensure that 
		there is no injection of kinetic helicity in the system.
	Finally, the last term--- where $\Omega$ is the rotation rate---mimics the Coriolis force and is made explicitly imaginary to ensure that it does not explicitly inject energy into the system.
	
	\begin{figure}
		\includegraphics[scale = 1]{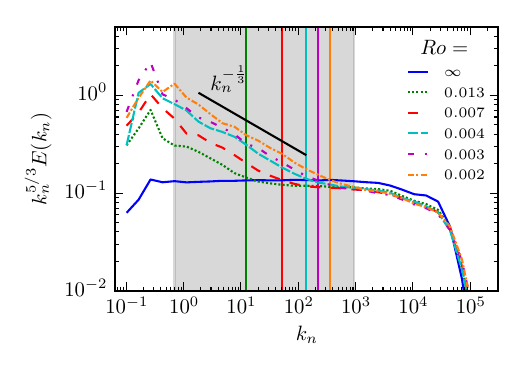}
		\caption{Loglog plots of the compensated energy spectrum
			$k_n^{5/3}E(k_n)$ versus the wavenumber $k_n$ for different Rossby
			numbers (see legend) from our simulations of the helical shell model.
			For $Ro = \infty$, the plateau (over several decades and shown by the
			shaded region) confirms the Kolmogorov scaling $E(k_n) \sim k_n^{-5/3}$
			for non-rotating, homogeneous and isotropic turbulence. As the Rossby
			number decreases, corresponding to an increased level of rotation, the
			compensated spectrum to the left of the Zeman scale (shown by
			vertical lines with colors corresponding to the respective $Ro$
			numbers) departs from the plateau with an additional scaling factor
			which asymptotes to $k_n^{-1/3}$, and hence $E(k_n) \sim k_n^{-2}$, as
			$Ro \ll 1$ and for wavenumbers lower than the Zeman wavenumber.}
		\label{fig:spectra}
	\end{figure}
	
	In our simulations, we use a total of $N = 32$ shells (with $k_0 =
	1/16$ and $\lambda = 1.62$), $\nu = 10^{-7}$ ($Re \sim 10^7$), and for
	time-marching an exponential fourth-order Runge-Kutta scheme, with a
	time-step $\delta t = 10^{-4}$, to factor in the stiffness of these
	coupled ordinary differential equations. We initialise our velocity
	field ($u_n^\pm = \sqrt{k_n} \exp(i\theta)$, for $n\le 4$ and $u_n^\pm
	= \sqrt{k_n} \exp(-k_n^2) \exp(i\theta)$  for $n \ge 5$, where $\theta
	\in [0,2\pi]$ is a random phase) and force the system  to a
	statistically steady state before turning on the Coriolis term.  
	
Analogous to the case of turbulence generated in a rotating fluid governed by
the Navier--Stokes equation, the Rossby number is defined as $Ro:=U_{\rm
rms}/\Omega L_{0}$, where $U_{\rm rms}= (\sum_{n} \mid u_n \mid ^ 2)^{1/2}$ is
the root-mean-square velocity and $L_{0}= 1/k_0$ is the integral length
scale~\cite{Biferale-review}. We perform simulations with several different
values of $Ro$; in this paper, for clarity, we present results mostly for the
cases $Ro = \infty$ (no rotation), $0.004, 0.003$, and
$0.002$ (corresponding to $\Omega = 0, 10,15$, and $20$, respectively).

	\section{Structure functions in rotating turbulence}
	\label{known_results}
	
	The Coriolis force plays a significant effect on the
	statistics of turbulence. Dimensionally, rotation sets an (inverse) time-scale
	in the problem resulting in a characteristic (Zeman) scale $l_\Omega \sim
	\sqrt{{\epsilon}/{{\Omega}^3}}$ (or wavenumber $k_\Omega \sim
	\sqrt{{{\Omega}^3}/{\epsilon}}$) where the rotational and fluid (eddy)
	turnover time-scales match~\cite{Zeman1994}.  For finitely small values of the Zeman scale
	(corresponding to $Ro\ll 1$), the two-point statistics, most usefully
	characterized by the (Fourier space) kinetic energy spectrum $E(k) = |u(k)|^2$,
	shows a dual-cascade~\cite{Zeman1994,Zhou1995,Yeung1998,Baroud2002,Baroud2003,Hattori2004,Biferale2016}: For wavenumbers $k < k_\Omega$, $E(k) \sim k^{-2}$ 
	and for $k > k_\Omega$, the usual Kolmogorov spectrum $E(k) \sim k^{-5/3}$.
	The dual cascade energy spectrum phenomenology is central to theories of
	rotating turbulence. It is tempting to now ask if low-dimensional dynamical
	systems, which mimic the formal structure of the Navier--Stokes equation but
	are neither rigorously derived from them nor sensitive to the geometrical
	reorganisation of flows under rotation, show any evidence of this new
	rotation-induced scaling regime.  Remarkably, simulations of the shell
	model---which is devoid of geometry but only respects the formal  structure of
	the Navier--Stokes equation---shows the exact same scaling behaviour. {(For the
	shell model, the energy spectrum is defined as $E(k_n) =  |u(k_n)|^2/k_n$ with an
	associated Zeman scale defined as above.)}
	
	A convenient way to see the point and extent of departure from the Kolmogorov
	$k^{-5/3}$ scaling is to look at plots of the compensated spectrum $k_n^{5/3}E(k_n)$
	versus $k_n$, for different strengths of the Coriolis force. In
	Fig.~\ref{fig:spectra}, we present representative plots of this compensated
	spectra for different values of the Rossby  number. For clarity, we show by vertical lines,
	the Zeman wavenumber corresponding to different values of $Ro$ and shade
	the inertial range which would have been present in the absence of rotation;
	for easy comparison we also show results from simulations of non-rotating, homogeneous, isotropic 
	turbulence corresponding to $Ro = \infty$.
	We immediately note the flatness of the compensated spectrum---before
	falling-off in the deep dissipation range---for all values of $Ro$ as long
	as $k_n > k_\Omega$. This is in sharp contrast to the steeper slopes of the
	spectrum, as evidenced by the departure from the plateau, for finite rotation ($Ro \neq \infty$)
	at scales  $k_n < k_\Omega$. In the limit of strong rotation  $Ro
	\ll 1$, the compensated spectrum reaches a slope $E(k_n) \sim k_n^{-1/3}$
	(indicated by the black-solid line), for $k_n < k_\Omega$, corresponding to
	prediction of the secondary scaling regime $E(k) \sim k^{-2}$ for wavenumbers
	smaller than the Zeman wavenumber. The dual-scaling behaviour observed in our spectrum is consistent 
	with those observed earlier in fully resolved direct numerical simulations 
	for the so-called isotropic energy spectrum~\cite{Biferale2016,Mininni2010,Yeung1998}, which is obtained by ignoring the rotation-induced 
	anisotropy and averaging over spherical Fourier shells; interestingly, earlier results also suggest that 
	the scaling of this isotropic spectrum is identical to the one obtained from just considering the wavevectors 
	perpendicular to the axis of rotation~\cite{Thiele2009,Mininni2010}.
	
	\begin{figure}
		\includegraphics[scale = 1]{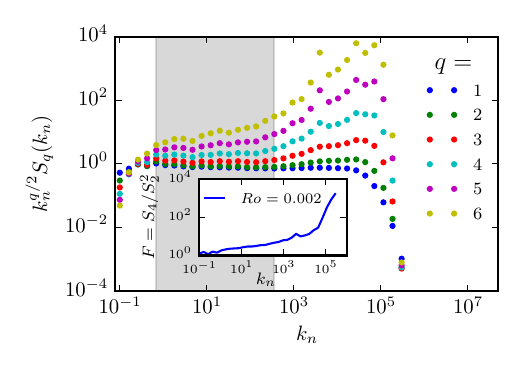}
		\caption{{Compensated plots of the equal-time structure functions (compensated by the dimensional 
		prediction $q/2$ for $Ro \to 0$) of various orders for $Ro = 0.002$. The shaded region marks the range 
		of inertial scales (lower than the Zeman wavenumber) where rotation modifies the scaling behaviour. The departure from 
		a plateau, especially for higher orders, shows the existence of intermittency corrections to the dimensional prediction.
		(Inset) A plot of the flatness versus the wave-number as a measure scale-by-scale 
		intermittency for the same value of the Rossby number.}}
		\label{fig:comp_Sq}
	\end{figure}

	The nature of the energy spectrum from our shell model is consistent with the phenomenology and dimensional predictions
	which ignore any corrections due to intermittency. Indeed it is well-known that
	intermittency corrections in the energy spectrum, that is essentially related
	to the second-order structure function through a Fourier transform, is
	notoriously hard to detect~\cite{Frisch1995}. Hence we must turn our attention to higher-order
	structure functions and calculate the scaling exponents $\xi_q$ (for $k_n <
	k_\Omega$), defined via $S_q(k_n) \equiv \langle |u_n|^q \rangle \sim k_n^{-\xi_q}$, 
	for different values of $Ro$. If we ignore the effects of intermittency, 
	we obtain $\xi_q = q/2$ as $Ro \to 0$ (strong rotation limit) and  $\zeta_q = 
	q/3$ as $Ro \to \infty$ (the familiar Kolmogorov result for non-rotating, homogeneous and isotropic turbulence).  
	Measurements \cite{Muller_2007,Mininni2009a}, as we also show below, suggest that for any
	finite Rossby number $\xi_q(\neq q/2)$ is a non-linear, concave function of $q$ (multiscaling stemming from the intermittency effects) and it is 
	only in the limit $Ro \to 0$, that the scaling exponents get close to the dimensional
	prediction. (Such a definition of structure functions in a shell model is consistent with  
	those defined for longitudinal velocity increments in the Navier--Stokes equation and DNSs; with the addition of rotation, as we see later in this paper, the shell model definition is consistent with the structure functions evaluated for the longitudinal velocity increments evaluated in the direction perpendicular to the rotation axis in direct numerical simulations. There is, of course, no rigorous way to prove this.)
	
	\begin{figure}
		\includegraphics[scale = 1]{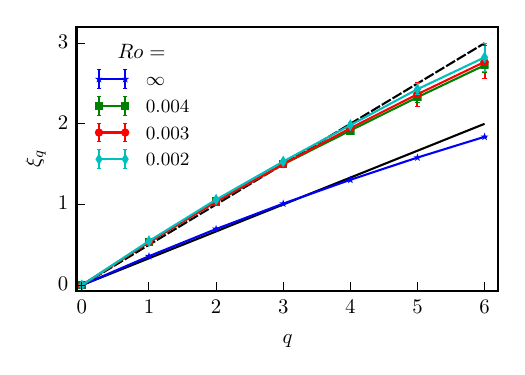}
		\caption{The scaling exponents $\xi_q$ (with error bars and connected
			by lines as a guide to the eye), from our shell model simulations, of
			the equal-time structure function for different Rossby numbers (see Table I and, for 
				comparison with {direct numerical simulation}s, Table II),
			including the case of no rotation ($Ro = \infty$).  For finite values
			of rotation, the exponents differ significantly from those obtained for
			non-rotating turbulence as well (lower set of data points, in blue,
			with the Kolmogorov scaling $q/3$ shown by a black solid line) as the
			dimensional prediction $\xi = q/2$ (black dashed line) for rotating
			flows. Nevertheless as $Ro \ll 1$, the exponents seem to asymptote to
			the dimensional prediction suggesting a suppression of intermittency in
			the flow and consequently a loss of multiscaling.} 
		\label{fig:xi}
	\end{figure}
	
	{In order to get a sense of how close the structure functions are to the dimensional prediction 
	$S_q(k_n) \sim k_n^{-q/2}$ (as $Ro \to 0$), we first adopt a strategy similar to that used more traditionally for the energy spectrum 
	(Fig.~\ref{fig:spectra}), namely to look at plots of the compensated structure function $k_n^{q/2}S_q(k_n)$ for a sufficiently 
	small value of the Rossby number. In Fig.~\ref{fig:comp_Sq}, we plot these compensated structure functions for several 
	orders for $Ro = 0.002$. The shading region marks the inertial range of scales smaller than the Zeman wavenumber. These plots 
	suggest that there is a clear departure from the dimensional prediction, especially for higher orders, due to the 
	familiar intermittency corrections. Indeed, in the inset of the same figure and for the same Rossby number, we plot the flatness 
	$S_4/S_2^2$ as a function of the wavenumber to illustrate more clearly the issue of scale-by-scale intermittency for such systems.}
	
	{All of these suggests that it is vital now to make our observations more quantitative by actually measuring 
	the equal-time exponents 
	for different values of the Rossby number. As is common to such measurements, we use the extended-self-similarity (ESS) procedure 
	to extract more reliable estimates of the scaling exponents~\cite{Benzi1993,Chakraborty2010a,Chakraborty2012}. 
	Furthermore, given the spacing of wavenumbers in a shell model a local slope 
	analysis, a natural choice for data from direct numerical simulations, is not as useful. We therefore use a different statistical 
	measure of the errors on our measurement. We obtain $S_q(k_n)$, and thus $\xi_q$, after averaging over several large eddy turn 
	over times. We then repeat this for 49 other statistically independent measurements to obtain 50 statistically independent 
	measurements of $\xi_q$. The mean of these yield the final exponents that we quote and their standard deviations become our error bars.}

	In Fig.~\ref{fig:xi}, we plot the equal-time scaling exponents $\xi_q$
	as a function of $q$ (also listed in Table I for different (small)
	values of $Ro$), from our shell model simulations, for different
	strengths of the Coriolis force.  For comparison, we also show the
	exponents $\xi_q$ for non-rotation turbulence ($\Omega = 0; Ro =
	\infty$) and the associated black solid line indicating the $q/3$
	prediction of Kolmogorov.  The black dashed line corresponds to the
	dimensional prediction $q/2$ and our measurements clearly show $\xi_q
	\neq q/2$, with the effect becoming more pronounced~\cite{Baroud2002,Baroud2003} for $q >
	3$.  For rotating flows, we of course
	present results in Fig.~\ref{fig:xi} for sufficiently low Rossby
	numbers. This is because for weak rotation (large Rossby numbers), the
	Zeman wavenumber is small leading to a shrinking inertial range (see
	Fig.~\ref{fig:spectra}) not amenable to the extraction of higher order
	scaling exponents. Hence, we do not show the exponents for such high
	Rossby numbers in Fig.~\ref{fig:xi}. However, as $Ro
	\to 0$, the scaling exponents tend to asymptote to values more
	consistent with the dimensional prediction showing a strong depletion
	of intermittency effects. This behaviour has been noted in  earlier
	studies~\cite{Muller_2007,Seiwert2008,Mininni2009a} that showed that
	strong rotation leads to a depletion of intermittency effects in
	turbulent flows. What is intriguing is that this feature is faithfully
	reproduced in a low-dimensional dynamical system---shell model---which
	is insensitive to any geometrical effects and the proliferation of
	columnar vortices in real flows or solutions of the Navier--Stokes
	equation.  Furthermore, the exponents from our shell model
	simulations (Fig.~\ref{fig:xi} and Table I) are in agreement  
	with those reported, for similar Rossby numbers, from
	direct numerical simulations of earlier studies,
	such as the work of Thiele and M\"uller~\cite{Thiele2009} (see Table
	II). {We note, in passing, that measurements of equal-time 
	exponents for wavenumbers $k > k_\Omega$ (but smaller than the dissipation range) 
	yield exponents which are consistent with what is known for non-rotating, homogeneous and isotropic, 
	three-dimensional turbulence.}

	\section{Energy Dissipation Rate}
	
	The three-dimensional Navier--Stokes equation is known to be invariant under
	suitable scaling transformations~\cite{Frisch1995} with a scaling exponent $h$
	which allows us to write the (scalar) velocity increments $\delta u_r$ across a
	scale $r$ as $\delta u_r \sim r^h$. Phenomenologically, the scale-dependent
	mean kinetic energy dissipation rate $\epsilon(r) \sim \frac{\delta u_r^3}{r}
	\sim r^{\alpha-1}$, where $\alpha = 3h$. For homogeneous and isotropic
	turbulence, Kolmogorov theory, in the absence of intermittency or multifractal
	statistics, predicts $h=1/3$ which ensures, in the inertial range, a
	scale-independent dissipation rate equal to the constant energy flux across
	these scales. Real turbulence, though, is multifractal.  Thus the dissipation
	field cannot be characterized by a unique choice of $\alpha$ but rather by
	its singularity spectrum $f(\alpha)$ and the mass function of Renyi dimension
	$\tau(q)$, both of which we define precisely later. 
	
	In a three-dimensional flow the local energy dissipation rate 
	\begin{equation}\label{def:epsilon}
		\varepsilon( \mb{x}) = \frac{\nu}{2}\sum_{i,j} (\partial_i u_j + \partial_j u_i)^2,
	\end{equation}
	is a function of all three spatial directions. For shell models, given its lack of spatial structure, obtaining the analogue
	of such a scalar dissipation field amenable to a multifractal analysis
	is less obvious. Let us nevertheless assume that the shell model describes the
	flow in a spatial domain of size $L$ and that the energy dissipation rate can
	be \textit{defined} at any spatial position $x \in [0,L]$. Thus, keeping in
	mind the Richardson picture of energy cascade, it is natural to assume that
	beginning with the largest eddy of size $\sim L$, an energy cascade is
	set up in the system such that each eddy in a given generation $m$ of the cascade breaks up
	into 2 to provide the eddies of the next generation.  This suggests a
	hierarchical transfer of energy, scale-by-scale, such that at each scale $m \in
	\{0, 1, 2, \cdots, K\}$, the number of eddies is $2^m$ and the typical size of each
	eddy is $l_m = L/2^m \sim 1/k_m$ (since the wave-numbers in a shell model
	correspond to the inverse of the spatial scales).  The smallest scales, set by
	$K$, correspond to the viscosity-dominated Kolmogorov scale of the flow. Let us
	now focus on the $i$-th eddy (of size $l_m = L/2^m \sim 1/k_m$) out of the
	$2^m$ eddies of generation $m$. Denoting the energy of this eddy by
	$E_i^{(m)}$, the total energy at scale $l_m$ must correspond to the
	kinetic-energy of the $m$-th shell in our shell model: $|u_m|^2 \equiv 2
	\sum_{i = 1}^{2^m} E_i^{(m)}$ with an associated energy density $E_i^{(m)}/l_m$ in the 
	$i$-th eddy. Thus we adapt the multifractal cascade ideas of Meneveau and Sreenivasan~\cite{Meneveau1987} 
	and construct it for our shell model. Furthermore, we choose different fractions $p  \in (0.5,0.9]$ of the energy distribution amongst 
	the daughter eddies and find that our results are qualitatively insensitive to the particular choice of $p$; in this paper we report results 
	for $p = 0.7$.

	With these definitions, following Lepreti \textit{et
		al.}~\cite{Lepreti2006}, 
	the kinetic energy density $e(x)$ at a spatial location $x \in [0,L]$ is given by 
	the contributions from eddies of all scales which have their imprints 
	on a specific point $x$: $e(x)=\sum_{m=0}^KE_{s_m(x)}^{(m)}/l_m$ where $s_m(x):=\ceil{(x-1)/l_m}$ 
	and thence the one-dimensional energy dissipation rate 
	\begin{equation}
		\varepsilon(x)=2\nu\sum_{m=0}^Kk_m^2E_{s_m(x)}^{(m)}/l_m
	\end{equation} 
	in the shell model. We refer to the reader to Lepreti \textit{et al.} (2006)~\cite{Lepreti2006} for a detailed 
	description on how $\epsilon(x)$ is evaluated in the shell model at any given instant of time from 
	the knowledge of the energy content of the eddies in the previous time step.
	In our calculations, we choose the largest length scale $L$ as the one associated with the forcing shell, i.e., $n = 3$, and ${K} = 23$ to obtain a grid resolution $L/2^{20}$. 
	Finally, to obtain reliable statistics, we extract the spatial distribution $\epsilon(x)$ from 100 different, statistically-independent velocity configurations in the steady state.

	\begin{figure}
		\includegraphics[]{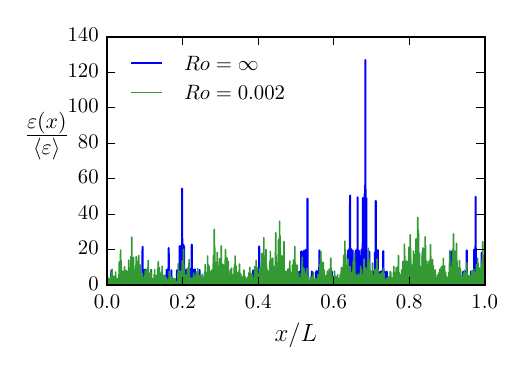}
		\caption{Reconstructed cuts of the energy dissipation rate following Lepreti \textit{et al.}~\cite{Lepreti2006} (see text) from our  
			simulations of the helical shell model. It suggests a suppression of intermittency when the flow is strongly rotating.}
		\label{fig:epsilon}
	\end{figure}
	
	In Fig.~\ref{fig:epsilon}, we show a representative plot of
	$\varepsilon(x)$, obtained from our shell model data as described above, for
	the case of the non-rotating flow and one with $Ro = 0.002$.
	Fig.~\ref{fig:epsilon} suggests that the behaviour of a spatial trace of
	the energy dissipation rate, at any given instant in time, obtained from the
	low-dimensional model is consistent with what is seen in direct numerical simulations of such flows (see also Fig.~\ref{fig:epsilondns} in Appendix~\ref{sec:DNS}, obtained from our 
	DNS). It is clear though that
	given the much higher Reynolds number that our shell model achieves (as well 
	as the more intermittent nature of such cascade models), the
	intensity of the intermittent peaks in the dissipation rate in
	Fig~\ref{fig:epsilon} are much stronger than what is seen in
	DNSs. 
	Furthermore, when we compare the cuts of
	these dissipation rates for the rotating and non-rotating cases, we do see a
	suggestion---the relatively \textit{calmer} traces of
	$\varepsilon$---that intermittency is suppressed (along with the degree of
	multifractality) as soon as we have a small enough Rossby number. However, this visual evidence 
	is hardly compelling and in order to substantiate our claim, we must resort to a 
	more quantitative characterization of this phenomenon through a full multifractal analysis.

	\section{Multifractal Analysis}

	\begin{figure*}
		\includegraphics[width=1.0\textwidth]{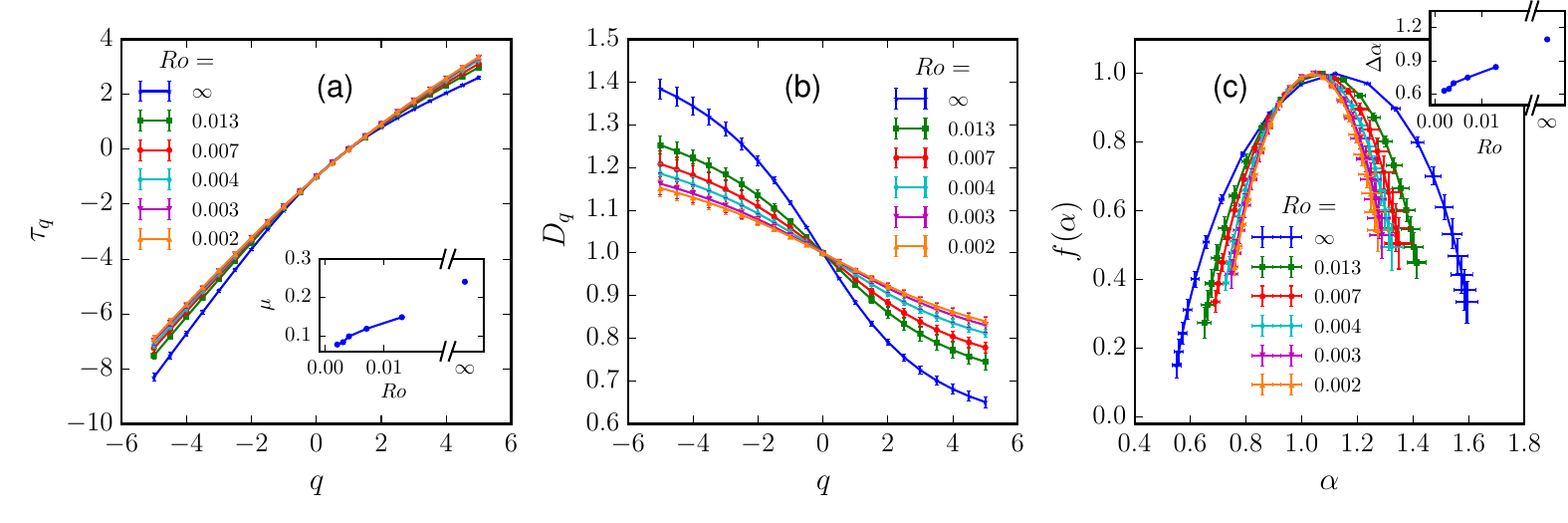}
		\caption{Signatures of decreasing intermittency with increased rotation. The figure depicts (a) the R\'enyi scaling exponents $\tau_q$ (inset: the intermittency exponent $\mu$ as a function of $Ro$), (b) the generalised dimension $D_q$, and (c)  the singularity spectrum $f(\alpha)$ (inset: widths $\Delta \alpha$ as a function of $Ro$), for different values of $Ro$, of the energy dissipation obtained from our simulations of the helical shell model and with $p = 0.7$ (see text).}
		\label{fig:mfa_SM}
	\end{figure*}	
	
	We set the stage for this analysis by defining, through the 1D cuts of the dissipation field $\varepsilon(x)$, a scale-dependent 
	energy dissipation $\varepsilon_r \equiv \int_{x \in I_r}\varepsilon(x)dx$ integrated in an interval $I_r$ of size $r$. If we choose the 
	interval all the way up to the integral scale $L$, this gives a reference scale-dependent dissipation $\varepsilon_L$ which allows us 
	to define the R\'enyi scaling exponent $\tau_q$ via 
	\begin{equation}
		\langle \epsilon_r^q\rangle \sim \epsilon_L^q \left(\frac{r}{L}\right)^{\tau_q}
		\label{tauq}
	\end{equation}
	for $L \gg r \gg \eta$. By using this exponent $\tau_q$, we can define the generalised dimension~\cite{Hentschel1983}
	$D_q = \tau_q / (q-1)$ and the multifractal singularity spectrum~\cite{Halsey1986,Ott2002} through a Legendre 
	transform of $\tau_q$ as $f(\alpha) = \min_q (q\alpha - \tau_q)$ where $\alpha = {d\tau_q}/{dq}$.
	As is traditional in this field, we characterize intermittency~\cite{Meneveau1991} through the exponent 
	$\mu = -({d^2 \tau_q}/{dq^2})_{q=0}$ ($\mu \approx 0.26$ for the non-rotating case) and the width of the singularity 
	spectrum $\Delta \alpha = \alpha_{\rm max} - \alpha_{\rm min}$, 
	where $\alpha_{\rm min}$ and $\alpha_{\rm max}$ are the strongest and weakest singularities, {respectively: Self-similar  
	solutions are naturally characterised by $\Delta \alpha = 0$ and $q$-independent generalised dimension.} 
	
	From our simulations of the
	helical shell model we first calculate $\tau_q$ via Eq.~\ref{tauq}, directly~\cite{Hentschel1983,Halsey1986} from the statistics of dissipation and from there the generalised dimension $D_q$ as well as the intermittency exponent $\mu$. 
	The singularity spectrum $f(\alpha)$ is obtained independently using the method proposed by Chhabra and Jensen~\cite{Chhabra1989a,Chhabra1989}; we have confirmed that a numerical calculation of 
	$f(\alpha)$ from $D_q$ by using the Legendre transform yields similar results but with larger error bars.
	In practice, we choose a long time series
	for the dissipation rate (Fig.~\ref{fig:epsilon}) from which we are
	able to compute $\tau_q$ for {$q\in [-5,5]$}. Such a large range of $q$ is
	important to ensure that the the generalised dimensions (computed through $\tau_q$) and the
	singularity spectrum shows definite signs of
	convergence: Namely, $D_q$ reaching a plateau for the largest $|q|$
	with an associated infinite slope of the singularity spectrum for the
	largest and smallest measured $\alpha$ (which corresponds
	to the extremal values of $D_q$ as $q \to -\infty$ and $q \to \infty$,
	respectively). Finally, we calculate $D_q$, $f(\alpha)$, and $\tau_q$
	(and thus the intermittency exponent $\mu$), from several independent
	runs; in Fig.~\ref{fig:mfa_SM} we plot the means of these measurements
	and show their standard deviations as error bars.

	In panel (a) of Fig.~\ref{fig:mfa_SM} we show plots of $\tau_q$ as a
	function of $q$ for different strengths of rotations. We notice, visually, a diminishing of 
	the curvature of $\tau_q$ around $q = 0$ suggesting a depletion of intermittency. This is confirmed 
	by measuring $\mu$ for different values of $Ro$ (inset, Fig.~\ref{fig:mfa_SM}(a)); as $Ro \to 0$, 
	there is a significant decrease in $\mu$ compared to the $Ro = \infty$ (non-rotating) value 
	for homogeneous and isotropic turbulence. This observations if further strengthened by looking 
	at plots of $D_q$ \textit{vs.} $q$ in Fig.~\ref{fig:mfa_SM}(b) which plateau (as $|q| \gg 1$) to values much closer  
	to 1 as the Rossby numbers decrease. This is confirmed in  
	in panel (c) where we show the singularity
	spectrum $f(\alpha)$ and the widths $\Delta\alpha$ as insets. With increasing
	rotation, the spectrum narrows, quantified by $\Delta\alpha$ which shows a
	monotonic decrease with $Ro$.
		
	Thus our multifractal analysis on the shell model data strongly suggests that the lack of 
	self-similarity (and intermittent behaviour) in the statistics of the dissipation rate---as measured through the generalised 
	dimension $D_q$, the singularity spectrum $f(\alpha)$ and  the exponent $\tau_q$---weakens considerably with increasing strengths of rotation. 
	Indeed in the limit $Ro \to 0$, it may be argued that the dissipation range statistics may well recover a truly self-similar form 
	which would be consistent with the dimensional (non-intermittent) prediction of the inertial range (anomalous) scaling exponent $\xi_q$.
	
	All of this then naturally brings us to an important question: For rotating turbulence (especially in the limit 
	$Ro \ll 1$), is there a way to bridge the statistics of the dissipation rates, characterised by generalised dimension $D_q$ with 
	the (anomalous) inertial range scaling exponents $\xi_q$? In other words, what would be the analogue of the familiar 
	$\zeta_q = q/3 + (q/3 - 1)(D_{q/3} - 1)$ result~\cite{Meneveau1987a} for non-rotating, homogeneous and isotropic turbulence?
	
	\begin{figure}
		\includegraphics[width=1.0\columnwidth]{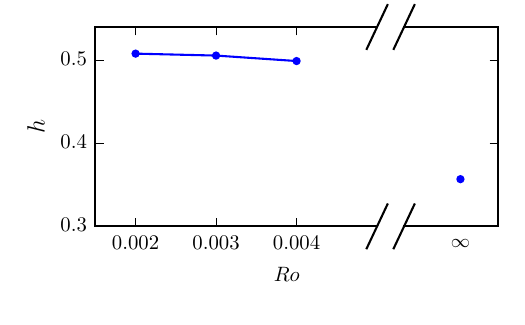}
		\caption{The scaling exponent $h$ versus the Rossby number $Ro$ from our shell model simulations. The exponent saturates to $h=1/2$ in the strong rotation limit $Ro \ll 1$ and to 
			$h = 1/3$  in the case of no rotation $Ro = \infty$.}
		\label{fig:h_vs_Omega}
	\end{figure}
	
	To answer this question, we make the following \textit{ansatz} when $Ro \ne \infty$:
	\begin{equation}
		\xi_q = hq + (hq - 1) (D_{hq} - 1)
	\end{equation}
	which can be re-arranged in the more useful form (for what is to follow):
	\begin{equation}
		\frac{\xi_q -1}{D_{h q}} + 1 = h q.
		\label{eqn:zetap_Dq}
	\end{equation}
	(For $Ro = \infty$ or $\Omega = 0$, the scaling exponent $h = 1/3$.) 
	
	By using our measurements of the scaling exponents $\xi_q$ and $D_q$, we use Eq.~(\ref{eqn:zetap_Dq}) 
	to obtain the scaling exponent $h$ as a function
	of $Ro$; our results are shown in Fig.~\ref{fig:h_vs_Omega}. Indeed, as we would expect from phenomenology (and also consistent with Fig.~\ref{fig:spectra}), as $Ro \to 0$, the exponent $h \to
	1/2$ while, in for the case of no rotation ($Ro = \infty$), $h = 1/3$. This leads us to conjecture, in the limit $Ro \to 0$, the
	following relation bridging the dissipation range and the inertial range
	statistics:
	
	\begin{equation}\label{eqn:result_eqn}
		\xi_q = q/2 + (q/2 - 1) (D_{q/2} - 1); \quad \quad Ro \to 0.
	\end{equation}
	
	{Furthermore, this relationship shows how the emergence
	of an approximate self-similarity ($\Delta \alpha \to 0$ and a
	$q$-independent generalised dimension) as $Ro \to 0$, leads to a
	progressively simple scaling (and not multi-scaling) of the exponents
	of the equal-time structure functions as shown in Fig.~\ref{fig:xi}.
	It is worth emphasizing that, in the limit $Ro \rightarrow 0$, the
	Zeman scale is pushed all the way up to the dissipation scale as
	indicated in Fig.~\ref{fig:spectra} making the K41 ($k^{-5/3}$) scaling
	no longer discernible: The $k^{-2}$ spectral scaling dominates all the
	way up to the dissipation scale. It is in this limit, of course, that
	our relation (\ref{eqn:result_eqn}) holds.}

		\begin{table}
			\begin{tabular*}{\linewidth}{@{\extracolsep{\fill}}| c | c | c | c |}
					\hline
					
					$Ro$ & $q$ & $\xi_q$ & $q/2 + (q/2-1)(D_{q/2} - 1)$ \\ [1mm]\hline
					
					& 1 & 0.513 $\pm$ 0.007 & 0.513 $\pm$ 0.001\\
					& 2 & 1 & 1.0 $\pm$ 0.0\\
			0.004	& 3 & 1.44 $\pm$ 0.03 & 1.463 $\pm$ 0.003	\\
					& 4 & 1.85 $\pm$ 0.05 & 1.905 $\pm$ 0.007	\\
					& 5 & 2.25 $\pm$ 0.07 & 2.31 $\pm$ 0.01	\\
					& 6 & 2.63 $\pm$ 0.09 & 2.70 $\pm$ 0.02	\\ \hline
					
					& 1 & 0.516 $ \pm $ 0.004 & 0.512 $\pm$ 0.002 \\
					& 2 & 1  & 1.0 $\pm$ 0.0 \\ 
			0.003	& 3 & 1.45  $ \pm $ 0.03  & 1.469 $\pm$ 0.004 \\
					& 4 & 1.88  $ \pm $ 0.07  & 1.92 $\pm$ 0.01 \\
					& 5 & 2.3   $ \pm $ 0.1   & 2.35 $\pm$ 0.02 \\
					& 6 & 2.7   $ \pm $ 0.2   & 2.77 $\pm$ 0.03 \\ \hline
					
					& 1 & 0.516 $\pm$ 0.008 & 0.510 $\pm$ 0.001 \\
					& 2 & 1  & 1.0 $\pm$ 0.0\\  
			0.002	& 3 & 1.45  $\pm$ 0.03  & 1.471 $\pm$ 0.003 \\
					& 4 & 1.87  $\pm$ 0.04  & 1.913 $\pm$ 0.007 \\
					& 5 & 2.29  $\pm$ 0.07  & 2.35 $\pm$ 0.01 \\
					& 6 & 2.7   $\pm$ 0.2   & 2.78 $\pm$ 0.02 \\ \hline
				\end{tabular*}
			\caption{{Summary of results from simulations of our shell model for different 
Rossby numbers.  The excellent agreement between columns 3 ($\xi_q$, the 
left hand side of Eq.~(\ref{eqn:result_eqn})) and 4 ($q/2 + (q/2-1)(D_{q/2} - 1)$, 
			the right hand side of Eq.~(\ref{eqn:result_eqn})) is evidence for the central theoretical prediction of this work.}}
			\label{table:validationSM}
		\end{table}

		Our theoretical prediction, while similar in spirit to earlier results of the homogeneous and isotropic turbulence, 
		still needs an independent check from numerical data. A convincing way to do this is to independently evaluate the left-hand-side ($\xi_q$, corresponding 
		to inertial range statistics) and the right-hand-side 
		($q/2 + (q/2 - 1) (D_{q/2} - 1)$, corresponding to dissipation range statistics) of Eq.~(\ref{eqn:result_eqn}) from our simulations 
		of the shell model and check if the equality holds for sufficiently small values of $Ro$ and several values of $q$. Given that we have already 
		calculated the equal-time exponents $\xi_q$ (Fig.~\ref{fig:xi}) and the generalised dimensions $D_q$ (Fig.~\ref{fig:mfa_SM}(b)) 
		for different values of $q$, it is straightforward now to carry out this validation. In Table I, we show values of the left-hand-side (column 3) 
		right-hand-side (column 4) of Eq.~(\ref{eqn:result_eqn}) for $q = 1, 2, ..., 6$; furthermore we present results for 3 different, but sufficiently small, 
		values of $Ro$. A comparison of columns 3 and 4, for any given $q$ (and $Ro$), leaves little doubt as to the validity of our prediction bridging 
		the inertial range and the dissipation range statistics.

	\begin{table}
		\begin{tabular*}{\linewidth}{@{\extracolsep{\fill}}| c | c | c | c |}
				\hline
				$Ro$	&  $q$ & $\xi_q$ & $q/2 + (q/2-1)(D_{q/2} - 1)$\\ [1mm] \hline
					& 1 & 0.52 $\pm$ 0.003   & 0.528 $\pm$ 0.006 \\
					& 2 & 1          & 1.0   		 $\pm$ 0.0 \\
			0.005	& 3 & 1.43 $\pm$ 0.006   & 1.42 $\pm$ 0.02 \\
					& 4 & 1.82  $\pm$ 0.02   & 1.80 $\pm$ 0.04 \\
					& 5 & 2.15  $\pm$ 0.03   & 2.13 $\pm$ 0.08 \\
					& 6 & 2.42 $\pm$ 0.05    & 2.4 $\pm$ 0.1 \\ \hline
					
					& 1 & 0.52 $\pm$ 0.002   & 0.522 $\pm$ 0.005 \\
					& 2 & 1          & 1.0   		 $\pm$  0.0\\      
			0.001	& 3 & 1.43 $\pm$ 0.007   & 1.43 $\pm$ 0.01 \\
					& 4 & 1.80  $\pm$ 0.02   & 1.82 $\pm$ 0.02 \\
					& 5 & 2.13  $\pm$ 0.03   & 2.17 $\pm$ 0.04 \\
					& 6 & 2.42 $\pm$ 0.05    & 2.48 $\pm$ 0.05 \\ \hline
			\end{tabular*}
			\caption{{A summary of results from direct numerical simulations confirming the validity of 
			Eq.~(\ref{eqn:result_eqn}) for two different values of the Rossby number (column 1). 
			The equal-time exponents $\xi_q$ (column 3) for the plane perpendicular to the axis of rotation, 
			are those reported, from the DNS studies by Thiele and M\"uller~\cite{Thiele2009};
				the generalised dimension $D_q$ (column 4) are 
				from our dissipation rate statistics for the plane perpendicular to the axis of 
				rotation (Fig.~\ref{fig:epsilondns}a).  
			The excellent agreement between columns 3 and 4 is compelling evidence for the validity of Eq.~(\ref{eqn:result_eqn}).}}
		\label{table:validationDNS}
	\end{table}
	

	Before we conclude, it is tempting to validate Eq.~(\ref{eqn:result_eqn}) against data from actual 
	direct numerical simulations. Although DNSs are not the central focus of this study, such an exercise is useful in this case for the sake of completeness. We use data for the equal-time exponents of the axis-perpendicular longitudinal velocity structure functions from the work of Thiele and M\"uller~\cite{Thiele2009} who used similarly small values of Rossby numbers. However in order to compute the generalised dimensions $D_q$ and thence the right-hand-side of Eq.~(\ref{eqn:result_eqn}), 
	we perform DNSs (see Appendix A for details) corresponding to the Rossby numbers used by Thiele and M\"uller~\cite{Thiele2009}. In a manner 
	exactly similar to Table I, we show, in Table II, the left-hand-side and right-hand-side of our bridge relation by using data from DNSs. Once again, 
	a comparison of columns 3 and 4 suggests clearly for small enough Rossby numbers, it is indeed possible to bridge the statistics 
	of the inertial and the dissipation ranges.

	\section{Conclusion}
	\label{conclusions}
	Equation~(\ref{eqn:result_eqn}) captures the key result in this study by
	bridging, in the limit of strongly rotating ($Ro \to 0$) turbulence,
	the statistics of the energy dissipation rate characterised by the
	generalised dimension $D_q$ with the scaling exponents $\xi_q$ of the
	moments of velocity increments over scales $r$ in the inertial range.
	Furthermore our analysis provides a more complete description of the
	depletion of intermittency in flows affected strongly by the Coriolis
	force.  Furthermore, our work shows that even for anisotropic system
	such as rotating turbulence, shell models, while being insensitive to
	the spatial reorganisation of the flow due to the Coriolis force, are
	nevertheless able to capture the scaling laws of two-point correlation
	functions in a way consistent with direct numerical simulations.
	Indeed, our work shows that following Lepreti \textit{et
	al.}~\cite{Lepreti2006} (see also Meneveau \textit{et
	al.}~\cite{Meneveau1987}), it is possible to extract a spatial profile
	of the dissipation rates in such shell models that can be used for
	future studies of the small-scale statistics of different forms of
	turbulence for whom such a low-dimensional dynamical systems
	representation exists. In this context, it is worth mentioning that there is another useful method~\cite{Jensen1999,Roux2004,Chakraborty2010,Chakraborty2010b} of generating a synthetic real space velocity field from the shell models; it could provide complementary insights in such studies.
	
	        Finally, we remind ourselves that this paper studies the problem of bridging inertial and dissipative 
		statistics in the limit where the mean helicity vanishes. Although we have some evidence that for small values of the mean helicity 
		our results and conclusions are unaltered, it would be important to study in future the problem 
		of non-zero helicity and how it should modify the central result captured in Eq.~(\ref{eqn:result_eqn}); the helical 
		shell model used in this paper is particularly useful from this point of view as it allows us to 
		introduce a mean helicity in a controlled way. Furthermore, Eq.~(\ref{eqn:result_eqn}) is an asymptotic result valid 
		in the vanishing Rossby number limit; we leave for future work possible Rossby number corrections to this prediction.
	
	\section*{Data Availability}
	The data that support the findings of this study are available from the corresponding author on request.
	\section*{Acknowledgments}
	SKR thanks Arijit Kundu for financial support through research grant no.~ECR/2018/001443 of the SERB (Govt. of India).	The simulations were performed on HPC2010, HPC2013, and Chaos clusters of IIT Kanpur, India.
	A part of this work was facilitated by a program organized at ICTS, Bengaluru: ``Summer
	Research Program on Dynamics of Complex Systems 2016'' (ICTS/Prog-dcs2016/06).
	SSR acknowledges support of the DAE, Govt.~of~India, under project no.~12-R\&D-TFR-5.10-1100 and DST (India) project MTR/2019/001553. The authors are grateful to Mahendra K. Verma for letting MKS use TARANG for the direct numerical simulations.

	\appendix
	\section{Direct Numerical Simulations}
	\label{sec:DNS}
\begin{figure}
		\includegraphics[]{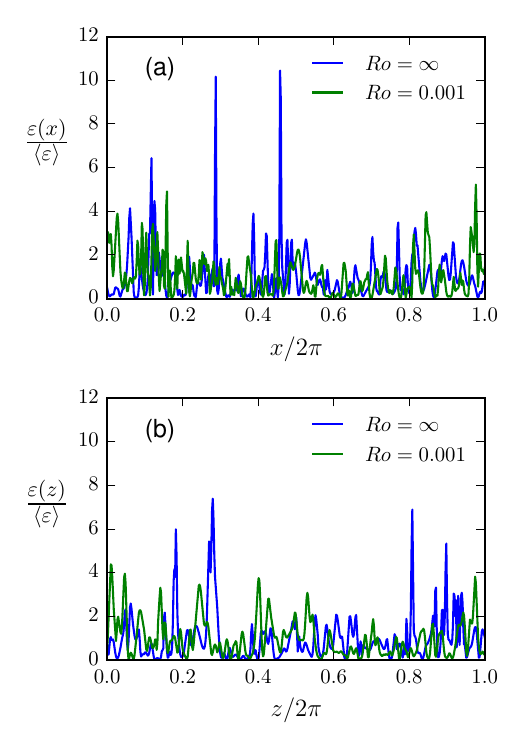}
		\caption{Representative snapshots of the one-dimensional cuts of the energy dissipation rates from our {DNS}s along the (a) x-axis and 
			(b) z-axis. A comparison of the panels with Fig.~\ref{fig:epsilon} show qualitatively similar features with a more intermittent---higher peaks---behaviour in the data from our shell model consistent with its much larger Reynolds number. These plots also suggest a suppression 
			of intermittency when the flow is strongly rotating.}
		\label{fig:epsilondns}
	\end{figure}

	Our direct numerical simulations, which were performed to provide additional support for our theoretical predictions, 
	were performed with a standard fully de-aliased
	pseudo-spectral method to solve Eq.~(\ref{eqn:NSE}) on a $2\pi$
	periodic cubic box, with $N^3 = 512^3$ collocation points, and a
	fourth-order Runge-Kutta scheme for time-marching. In our simulations, an 
	adaptive time-step $\delta t$   consistent with the
	Courant--Friedrich--Lewy (CFL) condition is employed and the viscosity $\nu =
	10^{-3}$ is chosen to obtain Taylor-scale based Reynolds numbers $Re_\lambda = 50$ and $58$, corresponding to rotation rates $\Omega = 8$ and $32$ 
	(or equivalently, $Ro=0.005$ and $0.001$), respectively. As is the case in the simulations of our shell model, we use a forcing
	
	\begin{eqnarray}
		\mathbf{{f}}(\mathbf{k}) =\frac{\epsilon {\mathbf{u}}(\mathbf{k}) }{n_f \left[{\mathbf{u}}(\mathbf{k})\cdot {\mathbf{u}}^*(\mathbf{k})\right]},
	\end{eqnarray}  

	which allows for a constant-energy injection (and no kinetic-helicity)
	at wavenumber $\mb{k}_f \in [{40,41}]$ ($n_f$ is the number of modes at
	this wavenumber) to drives the system to a statistically stationary,
	isotropic and homogeneous, turbulent regime.  Once we reach such a
	stationary state, we turn on the Coriolis force. We then wait a
	sufficiently long time for the rotating system to reach a statistically
	stationary state before measuring the local energy dissipation rates
	(Figs.~\ref{fig:epsilondns}(a) and (b)) and from there the generalised
	dimension $D_q$ necessary for the verification of Eq.(\ref{eqn:result_eqn}) (see Table II). 
	A convenient way to carry out a 
	multifractal analysis of energy dissipation field is to take several one-dimensional (1D) cuts of $\varepsilon( \mb{x})$ 
	parallel and perpendicular to the direction of rotation ($z-$axis). These 1D cuts 
	along the axis of rotation yield $\varepsilon(z) = \varepsilon(x_0, y_0, z)$; similarly, 
	for the plane perpendicular to the axis of rotation, we obtain
	$\varepsilon(x) = \varepsilon(x,y_0,z_0)$ and $\varepsilon(y) = \varepsilon(x_0,y,z_0)$. For reliable statistics, 
	we choose 49 cuts along each direction with different values of $x_0$, $y_0$ and $z_0 $ lying in the interval $[\pi/4, 3\pi/4 ]$. Considering that the data size for a cut is small, we employ a wavelet\cite{Meneveau1991a} leaders based method \cite{Jaffard2007,Wendt_2007} to calculate the R\'enyi scaling exponents, and thence the generalized dimensions.
	
	In Figs.~\ref{fig:epsilondns}(a) and (b) we show representative plots
	of one such cut for the reconstructed $\varepsilon(x)$ and
	$\varepsilon(z)$, respectively, normalised by the global mean, at a
	single instant of time for the non-rotating case and one with Rossby
	number $Ro = 0.001$ which illustrates that highly intermittent nature
	of the dissipation field persists even for such 1D cuts. It is worth remarking that 
	such fields are strikingly similar (with less extreme excursions) to the ones reported in the 
	paper from our shell model simulations. Finally, we mention in passing, that the singularity spectrum 
	$f(\alpha)$ and the exponent $\tau_q$ from these DNSs are in excellent agreement to those reported 
	in this paper from our shell model study.

	Our DNSs were performed by using the open-source code ``Tarang''~\cite{Verma2013,Chatterjee2018} developed at IIT Kanpur. 
	\section*{References}
	\bibliography{rathor_etal_manuscript}

\begin{thebibliography}{104}%
\makeatletter
\providecommand \@ifxundefined [1]{%
 \@ifx{#1\undefined}
}%
\providecommand \@ifnum [1]{%
 \ifnum #1\expandafter \@firstoftwo
 \else \expandafter \@secondoftwo
 \fi
}%
\providecommand \@ifx [1]{%
 \ifx #1\expandafter \@firstoftwo
 \else \expandafter \@secondoftwo
 \fi
}%
\providecommand \natexlab [1]{#1}%
\providecommand \enquote  [1]{``#1''}%
\providecommand \bibnamefont  [1]{#1}%
\providecommand \bibfnamefont [1]{#1}%
\providecommand \citenamefont [1]{#1}%
\providecommand \href@noop [0]{\@secondoftwo}%
\providecommand \href [0]{\begingroup \@sanitize@url \@href}%
\providecommand \@href[1]{\@@startlink{#1}\@@href}%
\providecommand \@@href[1]{\endgroup#1\@@endlink}%
\providecommand \@sanitize@url [0]{\catcode `\\12\catcode `\$12\catcode
  `\&12\catcode `\#12\catcode `\^12\catcode `\_12\catcode `\%12\relax}%
\providecommand \@@startlink[1]{}%
\providecommand \@@endlink[0]{}%
\providecommand \url  [0]{\begingroup\@sanitize@url \@url }%
\providecommand \@url [1]{\endgroup\@href {#1}{\urlprefix }}%
\providecommand \urlprefix  [0]{URL }%
\providecommand \Eprint [0]{\href }%
\providecommand \doibase [0]{http://dx.doi.org/}%
\providecommand \selectlanguage [0]{\@gobble}%
\providecommand \bibinfo  [0]{\@secondoftwo}%
\providecommand \bibfield  [0]{\@secondoftwo}%
\providecommand \translation [1]{[#1]}%
\providecommand \BibitemOpen [0]{}%
\providecommand \bibitemStop [0]{}%
\providecommand \bibitemNoStop [0]{.\EOS\space}%
\providecommand \EOS [0]{\spacefactor3000\relax}%
\providecommand \BibitemShut  [1]{\csname bibitem#1\endcsname}%
\let\auto@bib@innerbib\@empty
\bibitem [{\citenamefont {Frisch}(1995)}]{Frisch1995}%
  \BibitemOpen
  \bibfield  {author} {\bibinfo {author} {\bibfnamefont {U.}~\bibnamefont
  {Frisch}},\ }\href@noop {} {\emph {\bibinfo {title} {Turbulence: The Legacy
  of A. N. Kolmogorov}}}\ (\bibinfo  {publisher} {Cambridge University Press},\
  \bibinfo {year} {1995})\BibitemShut {NoStop}%
\bibitem [{\citenamefont {Lohse}\ and\ \citenamefont
  {Grossmann}(1993)}]{Lohse1993}%
  \BibitemOpen
  \bibfield  {author} {\bibinfo {author} {\bibfnamefont {D.}~\bibnamefont
  {Lohse}}\ and\ \bibinfo {author} {\bibfnamefont {S.}~\bibnamefont
  {Grossmann}},\ }\bibfield  {title} {\enquote {\bibinfo {title} {Intermittency
  in turbulence},}\ }\href@noop {} {\bibfield  {journal} {\bibinfo  {journal}
  {Physica A: Statistical Mechanics and its Applications}\ }\textbf {\bibinfo
  {volume} {194}},\ \bibinfo {pages} {519--531} (\bibinfo {year}
  {1993})}\BibitemShut {NoStop}%
\bibitem [{\citenamefont {Kolmogorov}(1941{\natexlab{a}})}]{Kolmogorov1941a}%
  \BibitemOpen
  \bibfield  {author} {\bibinfo {author} {\bibfnamefont {A.~N.}\ \bibnamefont
  {Kolmogorov}},\ }\bibfield  {title} {\enquote {\bibinfo {title} {{T}he local
  structure of turbulence in incompressible viscous fluid for very large
  {R}eynolds numbers},}\ }\href@noop {} {\bibfield  {journal} {\bibinfo
  {journal} {Dokl. Akad. Nauk SSSR}\ }\textbf {\bibinfo {volume} {30}},\
  \bibinfo {pages} {301--305} (\bibinfo {year}
  {1941}{\natexlab{a}})}\BibitemShut {NoStop}%
\bibitem [{\citenamefont {Kolmogorov}(1941{\natexlab{b}})}]{Kolmogorov1941b}%
  \BibitemOpen
  \bibfield  {author} {\bibinfo {author} {\bibfnamefont {A.~N.}\ \bibnamefont
  {Kolmogorov}},\ }\bibfield  {title} {\enquote {\bibinfo {title} {Dissipation
  of energy in locally isotropic turbulence},}\ }\href@noop {} {\bibfield
  {journal} {\bibinfo  {journal} {{D}okl. {A}kad. {N}auk SSSR}\ }\textbf
  {\bibinfo {volume} {32}},\ \bibinfo {pages} {16--18} (\bibinfo {year}
  {1941}{\natexlab{b}})}\BibitemShut {NoStop}%
\bibitem [{\citenamefont {Kolmogorov}(1962)}]{Kolmogorov1962}%
  \BibitemOpen
  \bibfield  {author} {\bibinfo {author} {\bibfnamefont {A.~N.}\ \bibnamefont
  {Kolmogorov}},\ }\bibfield  {title} {\enquote {\bibinfo {title} {A refinement
  of previous hypotheses concerning the local structure of turbulence in a
  viscous incompressible fluid at high reynolds number},}\ }\href@noop {}
  {\bibfield  {journal} {\bibinfo  {journal} {Journal of Fluid Mechanics}\
  }\textbf {\bibinfo {volume} {13}},\ \bibinfo {pages} {82--85} (\bibinfo
  {year} {1962})}\BibitemShut {NoStop}%
\bibitem [{\citenamefont {Kailasnath}, \citenamefont {Sreenivasan},\ and\
  \citenamefont {Stolovitzky}(1992)}]{Kailasnath-1992}%
  \BibitemOpen
  \bibfield  {author} {\bibinfo {author} {\bibfnamefont {P.}~\bibnamefont
  {Kailasnath}}, \bibinfo {author} {\bibfnamefont {K.}~\bibnamefont
  {Sreenivasan}}, \ and\ \bibinfo {author} {\bibfnamefont {G.}~\bibnamefont
  {Stolovitzky}},\ }\bibfield  {title} {\enquote {\bibinfo {title} {Probability
  density of velocity increments in turbulent flows},}\ }\href@noop {}
  {\bibfield  {journal} {\bibinfo  {journal} {Phys. Rev. Lett.}\ }\textbf
  {\bibinfo {volume} {68}},\ \bibinfo {pages} {2766--2769} (\bibinfo {year}
  {1992})}\BibitemShut {NoStop}%
\bibitem [{\citenamefont {La~Porta}\ \emph {et~al.}(2001)\citenamefont
  {La~Porta}, \citenamefont {Voth}, \citenamefont {Crawford}, \citenamefont
  {Alexander},\ and\ \citenamefont {Bodenschatz}}]{Bodenschatz-Nature}%
  \BibitemOpen
  \bibfield  {author} {\bibinfo {author} {\bibfnamefont {A.}~\bibnamefont
  {La~Porta}}, \bibinfo {author} {\bibfnamefont {G.~A.}\ \bibnamefont {Voth}},
  \bibinfo {author} {\bibfnamefont {A.~M.}\ \bibnamefont {Crawford}}, \bibinfo
  {author} {\bibfnamefont {J.}~\bibnamefont {Alexander}}, \ and\ \bibinfo
  {author} {\bibfnamefont {E.}~\bibnamefont {Bodenschatz}},\ }\bibfield
  {title} {\enquote {\bibinfo {title} {Fluid particle accelerations in fully
  developed turbulence},}\ }\href@noop {} {\bibfield  {journal} {\bibinfo
  {journal} {Nature}\ }\textbf {\bibinfo {volume} {409}},\ \bibinfo {pages}
  {1017--1019} (\bibinfo {year} {2001})}\BibitemShut {NoStop}%
\bibitem [{\citenamefont {Batchelor}\ and\ \citenamefont
  {Townsend}(1949)}]{Batchelor1949}%
  \BibitemOpen
  \bibfield  {author} {\bibinfo {author} {\bibfnamefont {G.~K.}\ \bibnamefont
  {Batchelor}}\ and\ \bibinfo {author} {\bibfnamefont {A.~A.}\ \bibnamefont
  {Townsend}},\ }\bibfield  {title} {\enquote {\bibinfo {title} {The nature of
  turbulent motion at large wave-numbers},}\ }\href@noop {} {\bibfield
  {journal} {\bibinfo  {journal} {Proceedings of the Royal Society of London.
  Series A. Mathematical and Physical Sciences}\ }\textbf {\bibinfo {volume}
  {199}},\ \bibinfo {pages} {238--255} (\bibinfo {year} {1949})}\BibitemShut
  {NoStop}%
\bibitem [{\citenamefont {Siggia}(1981)}]{Siggia1981}%
  \BibitemOpen
  \bibfield  {author} {\bibinfo {author} {\bibfnamefont {E.~D.}\ \bibnamefont
  {Siggia}},\ }\bibfield  {title} {\enquote {\bibinfo {title} {Numerical study
  of small-scale intermittency in three-dimensional turbulence},}\ }\href@noop
  {} {\bibfield  {journal} {\bibinfo  {journal} {Journal of Fluid Mechanics}\
  }\textbf {\bibinfo {volume} {107}},\ \bibinfo {pages} {375} (\bibinfo {year}
  {1981})}\BibitemShut {NoStop}%
\bibitem [{\citenamefont {Mandelbrot}(1972)}]{Mandelbrot1972}%
  \BibitemOpen
  \bibfield  {author} {\bibinfo {author} {\bibfnamefont {B.~B.}\ \bibnamefont
  {Mandelbrot}},\ }\bibfield  {title} {\enquote {\bibinfo {title} {Possible
  refinement of the lognormal hypothesis concerning the distribution of energy
  dissipation in intermittent turbulence},}\ }in\ \href@noop {} {\emph
  {\bibinfo {booktitle} {Statistical Models and Turbulence}}}\ (\bibinfo
  {publisher} {Springer Berlin Heidelberg},\ \bibinfo {year} {1972})\ pp.\
  \bibinfo {pages} {333--351}\BibitemShut {NoStop}%
\bibitem [{\citenamefont {Kraichnan}(1974)}]{Kraichnan1974}%
  \BibitemOpen
  \bibfield  {author} {\bibinfo {author} {\bibfnamefont {R.~H.}\ \bibnamefont
  {Kraichnan}},\ }\bibfield  {title} {\enquote {\bibinfo {title} {On
  kolmogorov's inertial-range theories},}\ }\href@noop {} {\bibfield  {journal}
  {\bibinfo  {journal} {Journal of Fluid Mechanics}\ }\textbf {\bibinfo
  {volume} {62}},\ \bibinfo {pages} {305} (\bibinfo {year} {1974})}\BibitemShut
  {NoStop}%
\bibitem [{\citenamefont {Mandelbrot}(1974)}]{Mandelbrot1974}%
  \BibitemOpen
  \bibfield  {author} {\bibinfo {author} {\bibfnamefont {B.~B.}\ \bibnamefont
  {Mandelbrot}},\ }\bibfield  {title} {\enquote {\bibinfo {title} {Intermittent
  turbulence in self-similar cascades: divergence of high moments and dimension
  of the carrier},}\ }\href@noop {} {\bibfield  {journal} {\bibinfo  {journal}
  {Journal of Fluid Mechanics}\ }\textbf {\bibinfo {volume} {62}},\ \bibinfo
  {pages} {331–358} (\bibinfo {year} {1974})}\BibitemShut {NoStop}%
\bibitem [{\citenamefont {Frisch}, \citenamefont {Sulem},\ and\ \citenamefont
  {Nelkin}(1978)}]{Frisch1978}%
  \BibitemOpen
  \bibfield  {author} {\bibinfo {author} {\bibfnamefont {U.}~\bibnamefont
  {Frisch}}, \bibinfo {author} {\bibfnamefont {P.-L.}\ \bibnamefont {Sulem}}, \
  and\ \bibinfo {author} {\bibfnamefont {M.}~\bibnamefont {Nelkin}},\
  }\bibfield  {title} {\enquote {\bibinfo {title} {A simple dynamical model of
  intermittent fully developed turbulence},}\ }\href@noop {} {\bibfield
  {journal} {\bibinfo  {journal} {Journal of Fluid Mechanics}\ }\textbf
  {\bibinfo {volume} {87}},\ \bibinfo {pages} {719--736} (\bibinfo {year}
  {1978})}\BibitemShut {NoStop}%
\bibitem [{\citenamefont {Frisch}\ and\ \citenamefont
  {Parisi}(1985)}]{Ghil1985}%
  \BibitemOpen
  \bibfield  {author} {\bibinfo {author} {\bibfnamefont {U.}~\bibnamefont
  {Frisch}}\ and\ \bibinfo {author} {\bibfnamefont {G.}~\bibnamefont
  {Parisi}},\ }\enquote {\bibinfo {title} {Turbulence and predictability in
  geophysical fluid dynamics and climate dynamics},}\ \ (\bibinfo  {publisher}
  {North-Holland Publ. Co., Amsterdam/New York},\ \bibinfo {year} {1985})\
  Chap.\ \bibinfo {chapter} {On the singularity structure of fully developed
  turbulence}, p.~\bibinfo {pages} {84}\BibitemShut {NoStop}%
\bibitem [{\citenamefont {Pandit}, \citenamefont {Perlekar},\ and\
  \citenamefont {Ray}(2009)}]{Pandit-Review}%
  \BibitemOpen
  \bibfield  {author} {\bibinfo {author} {\bibfnamefont {R.}~\bibnamefont
  {Pandit}}, \bibinfo {author} {\bibfnamefont {P.}~\bibnamefont {Perlekar}}, \
  and\ \bibinfo {author} {\bibfnamefont {S.~S.}\ \bibnamefont {Ray}},\
  }\bibfield  {title} {\enquote {\bibinfo {title} {Statistical properties of
  turbulence: An overview},}\ }\href@noop {} {\bibfield  {journal} {\bibinfo
  {journal} {Pramana}\ }\textbf {\bibinfo {volume} {73}},\ \bibinfo {pages}
  {157} (\bibinfo {year} {2009})}\BibitemShut {NoStop}%
\bibitem [{\citenamefont {Greenspan}(1968)}]{Greenspan1968}%
  \BibitemOpen
  \bibfield  {author} {\bibinfo {author} {\bibnamefont {Greenspan}},\
  }\href@noop {} {\emph {\bibinfo {title} {The Theory of Rotating Fluids}}}\
  (\bibinfo  {publisher} {Cambridge University Press},\ \bibinfo {year}
  {1968})\BibitemShut {NoStop}%
\bibitem [{\citenamefont {Moffatt}(1983)}]{Moffat_1983}%
  \BibitemOpen
  \bibfield  {author} {\bibinfo {author} {\bibfnamefont {H.~K.}\ \bibnamefont
  {Moffatt}},\ }\bibfield  {title} {\enquote {\bibinfo {title} {Transport
  effects associated with turbulence with particular attention to the influence
  of helicity},}\ }\href@noop {} {\bibfield  {journal} {\bibinfo  {journal}
  {Rep. Prog. Phys.}\ }\textbf {\bibinfo {volume} {46}},\ \bibinfo {pages}
  {621} (\bibinfo {year} {1983})}\BibitemShut {NoStop}%
\bibitem [{\citenamefont {Davidson}(2013)}]{davidson_2013}%
  \BibitemOpen
  \bibfield  {author} {\bibinfo {author} {\bibfnamefont {P.~A.}\ \bibnamefont
  {Davidson}},\ }\href@noop {} {\emph {\bibinfo {title} {Turbulence in
  Rotating, Stratified and Electrically Conducting Fluids}}}\ (\bibinfo
  {publisher} {Cambridge University Press},\ \bibinfo {year}
  {2013})\BibitemShut {NoStop}%
\bibitem [{\citenamefont {Barnes}(2001)}]{Barnes_2001}%
  \BibitemOpen
  \bibfield  {author} {\bibinfo {author} {\bibfnamefont {S.~A.}\ \bibnamefont
  {Barnes}},\ }\bibfield  {title} {\enquote {\bibinfo {title} {{An assessment
  of the rotation rates of the host stars of extrasolar planets}},}\
  }\href@noop {} {\bibfield  {journal} {\bibinfo  {journal} {Astrophys. J.}\
  }\textbf {\bibinfo {volume} {561}},\ \bibinfo {pages} {1095--1106} (\bibinfo
  {year} {2001})}\BibitemShut {NoStop}%
\bibitem [{\citenamefont {Cho}\ \emph {et~al.}(2008)\citenamefont {Cho},
  \citenamefont {Menou}, \citenamefont {Hansen},\ and\ \citenamefont
  {Seager}}]{James_2008}%
  \BibitemOpen
  \bibfield  {author} {\bibinfo {author} {\bibfnamefont {J.~Y.-K.}\
  \bibnamefont {Cho}}, \bibinfo {author} {\bibfnamefont {K.}~\bibnamefont
  {Menou}}, \bibinfo {author} {\bibfnamefont {B.~M.~S.}\ \bibnamefont
  {Hansen}}, \ and\ \bibinfo {author} {\bibfnamefont {S.}~\bibnamefont
  {Seager}},\ }\bibfield  {title} {\enquote {\bibinfo {title} {Atmospheric
  circulation of close-in extrasolar giant planets. i. global, barotropic,
  adiabatic simulations},}\ }\href@noop {} {\bibfield  {journal} {\bibinfo
  {journal} {Astrophysic. J.}\ }\textbf {\bibinfo {volume} {675}},\ \bibinfo
  {pages} {817} (\bibinfo {year} {2008})}\BibitemShut {NoStop}%
\bibitem [{\citenamefont {Le~Reun}\ \emph {et~al.}(2017)\citenamefont
  {Le~Reun}, \citenamefont {Favier}, \citenamefont {Barker},\ and\
  \citenamefont {Le~Bars}}]{Reun_2017}%
  \BibitemOpen
  \bibfield  {author} {\bibinfo {author} {\bibfnamefont {T.}~\bibnamefont
  {Le~Reun}}, \bibinfo {author} {\bibfnamefont {B.}~\bibnamefont {Favier}},
  \bibinfo {author} {\bibfnamefont {A.~J.}\ \bibnamefont {Barker}}, \ and\
  \bibinfo {author} {\bibfnamefont {M.}~\bibnamefont {Le~Bars}},\ }\bibfield
  {title} {\enquote {\bibinfo {title} {Inertial wave turbulence driven by
  elliptical instability},}\ }\href@noop {} {\bibfield  {journal} {\bibinfo
  {journal} {Phys. Rev. Lett.}\ }\textbf {\bibinfo {volume} {119}},\ \bibinfo
  {pages} {034502} (\bibinfo {year} {2017})}\BibitemShut {NoStop}%
\bibitem [{\citenamefont {Aurnou}\ \emph {et~al.}(2015)\citenamefont {Aurnou},
  \citenamefont {Calkins}, \citenamefont {Cheng}, \citenamefont {Julien},
  \citenamefont {King}, \citenamefont {Nieves}, \citenamefont {Soderlund},\
  and\ \citenamefont {Stellmach}}]{Aurnou_2015}%
  \BibitemOpen
  \bibfield  {author} {\bibinfo {author} {\bibfnamefont {J.}~\bibnamefont
  {Aurnou}}, \bibinfo {author} {\bibfnamefont {M.}~\bibnamefont {Calkins}},
  \bibinfo {author} {\bibfnamefont {J.}~\bibnamefont {Cheng}}, \bibinfo
  {author} {\bibfnamefont {K.}~\bibnamefont {Julien}}, \bibinfo {author}
  {\bibfnamefont {E.}~\bibnamefont {King}}, \bibinfo {author} {\bibfnamefont
  {D.}~\bibnamefont {Nieves}}, \bibinfo {author} {\bibfnamefont
  {K.}~\bibnamefont {Soderlund}}, \ and\ \bibinfo {author} {\bibfnamefont
  {S.}~\bibnamefont {Stellmach}},\ }\bibfield  {title} {\enquote {\bibinfo
  {title} {Rotating convective turbulence in earth and planetary cores},}\
  }\href@noop {} {\bibfield  {journal} {\bibinfo  {journal} {Physics of the
  Earth and Planetary Interiors}\ }\textbf {\bibinfo {volume} {246}},\ \bibinfo
  {pages} {52 -- 71} (\bibinfo {year} {2015})}\BibitemShut {NoStop}%
\bibitem [{\citenamefont {Smith}\ and\ \citenamefont
  {Waleffe}(1999)}]{Smith_1999}%
  \BibitemOpen
  \bibfield  {author} {\bibinfo {author} {\bibfnamefont {L.~M.}\ \bibnamefont
  {Smith}}\ and\ \bibinfo {author} {\bibfnamefont {F.}~\bibnamefont
  {Waleffe}},\ }\bibfield  {title} {\enquote {\bibinfo {title} {Transfer of
  energy to two-dimensional large scales in forced, rotating three-dimensional
  turbulence},}\ }\href@noop {} {\bibfield  {journal} {\bibinfo  {journal}
  {Phys. Fluids}\ }\textbf {\bibinfo {volume} {11}},\ \bibinfo {pages}
  {1608--1622} (\bibinfo {year} {1999})}\BibitemShut {NoStop}%
\bibitem [{\citenamefont {M\"{u}ller}\ and\ \citenamefont
  {Thiele}(2007)}]{Muller_2007}%
  \BibitemOpen
  \bibfield  {author} {\bibinfo {author} {\bibfnamefont {W.-C.}\ \bibnamefont
  {M\"{u}ller}}\ and\ \bibinfo {author} {\bibfnamefont {M.}~\bibnamefont
  {Thiele}},\ }\bibfield  {title} {\enquote {\bibinfo {title} {Scaling and
  energy transfer in rotating turbulence},}\ }\href@noop {} {\bibfield
  {journal} {\bibinfo  {journal} {EPL (Europhysics Letters)}\ }\textbf
  {\bibinfo {volume} {77}},\ \bibinfo {pages} {34003} (\bibinfo {year}
  {2007})}\BibitemShut {NoStop}%
\bibitem [{\citenamefont {Mininni}, \citenamefont {Alexakis},\ and\
  \citenamefont {Pouquet}(2009)}]{Mininni2009a}%
  \BibitemOpen
  \bibfield  {author} {\bibinfo {author} {\bibfnamefont {P.~D.}\ \bibnamefont
  {Mininni}}, \bibinfo {author} {\bibfnamefont {A.}~\bibnamefont {Alexakis}}, \
  and\ \bibinfo {author} {\bibfnamefont {A.}~\bibnamefont {Pouquet}},\
  }\bibfield  {title} {\enquote {\bibinfo {title} {Scale interactions and
  scaling laws in rotating flows at moderate rossby numbers and large reynolds
  numbers},}\ }\href@noop {} {\bibfield  {journal} {\bibinfo  {journal}
  {Physics of Fluids}\ }\textbf {\bibinfo {volume} {21}},\ \bibinfo {pages}
  {015108} (\bibinfo {year} {2009})}\BibitemShut {NoStop}%
\bibitem [{\citenamefont {Bartello}(1995)}]{Bartello_1995}%
  \BibitemOpen
  \bibfield  {author} {\bibinfo {author} {\bibfnamefont {P.}~\bibnamefont
  {Bartello}},\ }\bibfield  {title} {\enquote {\bibinfo {title} {Geostrophic
  adjustment and inverse cascades in rotating stratified turbulence},}\
  }\href@noop {} {\bibfield  {journal} {\bibinfo  {journal} {Journal of the
  Atmospheric Sciences}\ }\textbf {\bibinfo {volume} {52}},\ \bibinfo {pages}
  {4410--4428} (\bibinfo {year} {1995})}\BibitemShut {NoStop}%
\bibitem [{\citenamefont {M\'{e}tais}\ \emph {et~al.}(1996)\citenamefont
  {M\'{e}tais}, \citenamefont {Bartello}, \citenamefont {Garnier},
  \citenamefont {Riley},\ and\ \citenamefont {Lesieur}}]{metais_1996}%
  \BibitemOpen
  \bibfield  {author} {\bibinfo {author} {\bibfnamefont {O.}~\bibnamefont
  {M\'{e}tais}}, \bibinfo {author} {\bibfnamefont {P.}~\bibnamefont
  {Bartello}}, \bibinfo {author} {\bibfnamefont {E.}~\bibnamefont {Garnier}},
  \bibinfo {author} {\bibfnamefont {J.}~\bibnamefont {Riley}}, \ and\ \bibinfo
  {author} {\bibfnamefont {M.}~\bibnamefont {Lesieur}},\ }\bibfield  {title}
  {\enquote {\bibinfo {title} {Inverse cascade in stably stratified rotating
  turbulence},}\ }\href@noop {} {\bibfield  {journal} {\bibinfo  {journal}
  {Dynamics of Atmospheres and Oceans}\ }\textbf {\bibinfo {volume} {23}},\
  \bibinfo {pages} {193 -- 203} (\bibinfo {year} {1996})},\ \bibinfo {note}
  {stratified flows}\BibitemShut {NoStop}%
\bibitem [{\citenamefont {Yarom}, \citenamefont {Vardi},\ and\ \citenamefont
  {Sharon}(2013)}]{Yarom_2013}%
  \BibitemOpen
  \bibfield  {author} {\bibinfo {author} {\bibfnamefont {E.}~\bibnamefont
  {Yarom}}, \bibinfo {author} {\bibfnamefont {Y.}~\bibnamefont {Vardi}}, \ and\
  \bibinfo {author} {\bibfnamefont {E.}~\bibnamefont {Sharon}},\ }\bibfield
  {title} {\enquote {\bibinfo {title} {Experimental quantification of inverse
  energy cascade in deep rotating turbulence},}\ }\href@noop {} {\bibfield
  {journal} {\bibinfo  {journal} {Physics of Fluids}\ }\textbf {\bibinfo
  {volume} {25}},\ \bibinfo {pages} {085105} (\bibinfo {year}
  {2013})}\BibitemShut {NoStop}%
\bibitem [{\citenamefont {Bartello}, \citenamefont {M\'{e}tais},\ and\
  \citenamefont {Lesieur}(1994)}]{bartello_1994}%
  \BibitemOpen
  \bibfield  {author} {\bibinfo {author} {\bibfnamefont {P.}~\bibnamefont
  {Bartello}}, \bibinfo {author} {\bibfnamefont {O.}~\bibnamefont
  {M\'{e}tais}}, \ and\ \bibinfo {author} {\bibfnamefont {M.}~\bibnamefont
  {Lesieur}},\ }\bibfield  {title} {\enquote {\bibinfo {title} {Coherent
  structures in rotating three-dimensional turbulence},}\ }\href@noop {}
  {\bibfield  {journal} {\bibinfo  {journal} {Journal of Fluid Mechanics}\
  }\textbf {\bibinfo {volume} {273}},\ \bibinfo {pages} {1–29} (\bibinfo
  {year} {1994})}\BibitemShut {NoStop}%
\bibitem [{\citenamefont {Godeferd}\ and\ \citenamefont
  {Lollini}(1999)}]{godeferd_1999}%
  \BibitemOpen
  \bibfield  {author} {\bibinfo {author} {\bibfnamefont {F.~S.}\ \bibnamefont
  {Godeferd}}\ and\ \bibinfo {author} {\bibfnamefont {L.}~\bibnamefont
  {Lollini}},\ }\bibfield  {title} {\enquote {\bibinfo {title} {Direct
  numerical simulations of turbulence with confinement and rotation},}\
  }\href@noop {} {\bibfield  {journal} {\bibinfo  {journal} {J. Fluid Mech.}\
  }\textbf {\bibinfo {volume} {393}},\ \bibinfo {pages} {257–308} (\bibinfo
  {year} {1999})}\BibitemShut {NoStop}%
\bibitem [{\citenamefont {Zeman}(1994)}]{Zeman1994}%
  \BibitemOpen
  \bibfield  {author} {\bibinfo {author} {\bibfnamefont {O.}~\bibnamefont
  {Zeman}},\ }\bibfield  {title} {\enquote {\bibinfo {title} {A note on the
  spectra and decay of rotating homogeneous turbulence},}\ }\href@noop {}
  {\bibfield  {journal} {\bibinfo  {journal} {Physics of Fluids}\ }\textbf
  {\bibinfo {volume} {6}},\ \bibinfo {pages} {3221--3223} (\bibinfo {year}
  {1994})}\BibitemShut {NoStop}%
\bibitem [{\citenamefont {Hattori}, \citenamefont {Rubinstein},\ and\
  \citenamefont {Ishizawa}(2004)}]{Hattori2004}%
  \BibitemOpen
  \bibfield  {author} {\bibinfo {author} {\bibfnamefont {Y.}~\bibnamefont
  {Hattori}}, \bibinfo {author} {\bibfnamefont {R.}~\bibnamefont {Rubinstein}},
  \ and\ \bibinfo {author} {\bibfnamefont {A.}~\bibnamefont {Ishizawa}},\
  }\bibfield  {title} {\enquote {\bibinfo {title} {Shell model for rotating
  turbulence},}\ }\href@noop {} {\bibfield  {journal} {\bibinfo  {journal}
  {Physical Review E}\ }\textbf {\bibinfo {volume} {70}} (\bibinfo {year}
  {2004})}\BibitemShut {NoStop}%
\bibitem [{\citenamefont {Sreenivasan}\ and\ \citenamefont
  {Davidson}(2008)}]{sreenivasan_2008}%
  \BibitemOpen
  \bibfield  {author} {\bibinfo {author} {\bibfnamefont {B.}~\bibnamefont
  {Sreenivasan}}\ and\ \bibinfo {author} {\bibfnamefont {P.~A.}\ \bibnamefont
  {Davidson}},\ }\bibfield  {title} {\enquote {\bibinfo {title} {On the
  formation of cyclones and anticyclones in a rotating fluid},}\ }\href@noop {}
  {\bibfield  {journal} {\bibinfo  {journal} {Phys. Fluids}\ }\textbf {\bibinfo
  {volume} {20}},\ \bibinfo {pages} {085104} (\bibinfo {year}
  {2008})}\BibitemShut {NoStop}%
\bibitem [{\citenamefont {Pouquet}\ and\ \citenamefont
  {Mininni}(2010)}]{Pouquet_2010}%
  \BibitemOpen
  \bibfield  {author} {\bibinfo {author} {\bibfnamefont {A.}~\bibnamefont
  {Pouquet}}\ and\ \bibinfo {author} {\bibfnamefont {P.~D.}\ \bibnamefont
  {Mininni}},\ }\bibfield  {title} {\enquote {\bibinfo {title} {The interplay
  between helicity and rotation in turbulence: implications for scaling laws
  and small-scale dynamics},}\ }\href@noop {} {\bibfield  {journal} {\bibinfo
  {journal} {Philos. Trans. R. Soc. Lond., A}\ }\textbf {\bibinfo {volume}
  {368}},\ \bibinfo {pages} {1635--1662} (\bibinfo {year} {2010})}\BibitemShut
  {NoStop}%
\bibitem [{\citenamefont {Castello}\ and\ \citenamefont
  {Clercx}(2011)}]{Castello_2011}%
  \BibitemOpen
  \bibfield  {author} {\bibinfo {author} {\bibfnamefont {L.~D.}\ \bibnamefont
  {Castello}}\ and\ \bibinfo {author} {\bibfnamefont {H.~J.~H.}\ \bibnamefont
  {Clercx}},\ }\bibfield  {title} {\enquote {\bibinfo {title} {Lagrangian
  velocity and acceleration auto-correlations in rotating turbulence},}\
  }\href@noop {} {\bibfield  {journal} {\bibinfo  {journal} {Journal of
  Physics: Conference Series}\ }\textbf {\bibinfo {volume} {318}},\ \bibinfo
  {pages} {052028} (\bibinfo {year} {2011})}\BibitemShut {NoStop}%
\bibitem [{\citenamefont {Biferale}\ \emph {et~al.}(2016)\citenamefont
  {Biferale}, \citenamefont {Bonaccorso}, \citenamefont {Mazzitelli},
  \citenamefont {van Hinsberg}, \citenamefont {Lanotte}, \citenamefont
  {Musacchio}, \citenamefont {Perlekar},\ and\ \citenamefont
  {Toschi}}]{Biferale2016}%
  \BibitemOpen
  \bibfield  {author} {\bibinfo {author} {\bibfnamefont {L.}~\bibnamefont
  {Biferale}}, \bibinfo {author} {\bibfnamefont {F.}~\bibnamefont
  {Bonaccorso}}, \bibinfo {author} {\bibfnamefont {I.}~\bibnamefont
  {Mazzitelli}}, \bibinfo {author} {\bibfnamefont {M.}~\bibnamefont {van
  Hinsberg}}, \bibinfo {author} {\bibfnamefont {A.}~\bibnamefont {Lanotte}},
  \bibinfo {author} {\bibfnamefont {S.}~\bibnamefont {Musacchio}}, \bibinfo
  {author} {\bibfnamefont {P.}~\bibnamefont {Perlekar}}, \ and\ \bibinfo
  {author} {\bibfnamefont {F.}~\bibnamefont {Toschi}},\ }\bibfield  {title}
  {\enquote {\bibinfo {title} {Coherent structures and extreme events in
  rotating multiphase turbulent flows},}\ }\href@noop {} {\bibfield  {journal}
  {\bibinfo  {journal} {Physical Review X}\ }\textbf {\bibinfo {volume} {6}}
  (\bibinfo {year} {2016})}\BibitemShut {NoStop}%
\bibitem [{\citenamefont {Maity}, \citenamefont {Govindarajan},\ and\
  \citenamefont {Ray}(2019)}]{Maity}%
  \BibitemOpen
  \bibfield  {author} {\bibinfo {author} {\bibfnamefont {P.}~\bibnamefont
  {Maity}}, \bibinfo {author} {\bibfnamefont {R.}~\bibnamefont {Govindarajan}},
  \ and\ \bibinfo {author} {\bibfnamefont {S.~S.}\ \bibnamefont {Ray}},\
  }\bibfield  {title} {\enquote {\bibinfo {title} {Statistics of lagrangian
  trajectories in a rotating turbulent flow},}\ }\href@noop {} {\bibfield
  {journal} {\bibinfo  {journal} {Phys. Rev. E}\ }\textbf {\bibinfo {volume}
  {100}},\ \bibinfo {pages} {043110} (\bibinfo {year} {2019})}\BibitemShut
  {NoStop}%
\bibitem [{\citenamefont {Ibbetson}\ and\ \citenamefont
  {Tritton}(1975)}]{ibbetson_tritton_1975}%
  \BibitemOpen
  \bibfield  {author} {\bibinfo {author} {\bibfnamefont {A.}~\bibnamefont
  {Ibbetson}}\ and\ \bibinfo {author} {\bibfnamefont {D.~J.}\ \bibnamefont
  {Tritton}},\ }\bibfield  {title} {\enquote {\bibinfo {title} {Experiments on
  turbulence in a rotating fluid},}\ }\href@noop {} {\bibfield  {journal}
  {\bibinfo  {journal} {J.Fluid Mech.}\ }\textbf {\bibinfo {volume} {68}},\
  \bibinfo {pages} {639–672} (\bibinfo {year} {1975})}\BibitemShut {NoStop}%
\bibitem [{\citenamefont {Hopfinger}, \citenamefont {Browand},\ and\
  \citenamefont {Gagne}(1982)}]{hopfinger_1982}%
  \BibitemOpen
  \bibfield  {author} {\bibinfo {author} {\bibfnamefont {E.~J.}\ \bibnamefont
  {Hopfinger}}, \bibinfo {author} {\bibfnamefont {F.~K.}\ \bibnamefont
  {Browand}}, \ and\ \bibinfo {author} {\bibfnamefont {Y.}~\bibnamefont
  {Gagne}},\ }\bibfield  {title} {\enquote {\bibinfo {title} {Turbulence and
  waves in a rotating tank},}\ }\href@noop {} {\bibfield  {journal} {\bibinfo
  {journal} {J. Fluid. Mech.}\ }\textbf {\bibinfo {volume} {125}},\ \bibinfo
  {pages} {505–534} (\bibinfo {year} {1982})}\BibitemShut {NoStop}%
\bibitem [{\citenamefont {Bewley}\ \emph {et~al.}(2007)\citenamefont {Bewley},
  \citenamefont {Lathrop}, \citenamefont {Maas},\ and\ \citenamefont
  {Sreenivasan}}]{Bewley_2007}%
  \BibitemOpen
  \bibfield  {author} {\bibinfo {author} {\bibfnamefont {G.~P.}\ \bibnamefont
  {Bewley}}, \bibinfo {author} {\bibfnamefont {D.~P.}\ \bibnamefont {Lathrop}},
  \bibinfo {author} {\bibfnamefont {L.~R.~M.}\ \bibnamefont {Maas}}, \ and\
  \bibinfo {author} {\bibfnamefont {K.~R.}\ \bibnamefont {Sreenivasan}},\
  }\bibfield  {title} {\enquote {\bibinfo {title} {Inertial waves in rotating
  grid turbulence},}\ }\href@noop {} {\bibfield  {journal} {\bibinfo  {journal}
  {Physics of Fluids}\ }\textbf {\bibinfo {volume} {19}},\ \bibinfo {pages}
  {071701} (\bibinfo {year} {2007})}\BibitemShut {NoStop}%
\bibitem [{\citenamefont {Morize}, \citenamefont {Moisy},\ and\ \citenamefont
  {Rabaud}(2005)}]{Morize_2005}%
  \BibitemOpen
  \bibfield  {author} {\bibinfo {author} {\bibfnamefont {C.}~\bibnamefont
  {Morize}}, \bibinfo {author} {\bibfnamefont {F.}~\bibnamefont {Moisy}}, \
  and\ \bibinfo {author} {\bibfnamefont {M.}~\bibnamefont {Rabaud}},\
  }\bibfield  {title} {\enquote {\bibinfo {title} {Decaying grid-generated
  turbulence in a rotating tank},}\ }\href@noop {} {\bibfield  {journal}
  {\bibinfo  {journal} {Phys. Fluids}\ }\textbf {\bibinfo {volume} {17}},\
  \bibinfo {pages} {095105} (\bibinfo {year} {2005})}\BibitemShut {NoStop}%
\bibitem [{\citenamefont {Moisy}\ \emph {et~al.}(2011)\citenamefont {Moisy},
  \citenamefont {Morize}, \citenamefont {Rabaud},\ and\ \citenamefont
  {Sommeria}}]{moisy_2011}%
  \BibitemOpen
  \bibfield  {author} {\bibinfo {author} {\bibfnamefont {F.}~\bibnamefont
  {Moisy}}, \bibinfo {author} {\bibfnamefont {C.}~\bibnamefont {Morize}},
  \bibinfo {author} {\bibfnamefont {M.}~\bibnamefont {Rabaud}}, \ and\ \bibinfo
  {author} {\bibfnamefont {J.}~\bibnamefont {Sommeria}},\ }\bibfield  {title}
  {\enquote {\bibinfo {title} {Decay laws, anisotropy and cyclone–anticyclone
  asymmetry in decaying rotating turbulence},}\ }\href@noop {} {\bibfield
  {journal} {\bibinfo  {journal} {Journal of Fluid Mechanics}\ }\textbf
  {\bibinfo {volume} {666}},\ \bibinfo {pages} {5–35} (\bibinfo {year}
  {2011})}\BibitemShut {NoStop}%
\bibitem [{\citenamefont {Yarom}, \citenamefont {Salhov},\ and\ \citenamefont
  {Sharon}(2017)}]{Yarom_2017}%
  \BibitemOpen
  \bibfield  {author} {\bibinfo {author} {\bibfnamefont {E.}~\bibnamefont
  {Yarom}}, \bibinfo {author} {\bibfnamefont {A.}~\bibnamefont {Salhov}}, \
  and\ \bibinfo {author} {\bibfnamefont {E.}~\bibnamefont {Sharon}},\
  }\bibfield  {title} {\enquote {\bibinfo {title} {Experimental quantification
  of nonlinear time scales in inertial wave rotating turbulence},}\ }\href@noop
  {} {\bibfield  {journal} {\bibinfo  {journal} {Phys. Rev. Fluids}\ }\textbf
  {\bibinfo {volume} {2}},\ \bibinfo {pages} {122601} (\bibinfo {year}
  {2017})}\BibitemShut {NoStop}%
\bibitem [{\citenamefont {Sharma}\ \emph {et~al.}(2018)\citenamefont {Sharma},
  \citenamefont {Kumar}, \citenamefont {Verma},\ and\ \citenamefont
  {Chakraborty}}]{Sharma2018a}%
  \BibitemOpen
  \bibfield  {author} {\bibinfo {author} {\bibfnamefont {M.~K.}\ \bibnamefont
  {Sharma}}, \bibinfo {author} {\bibfnamefont {A.}~\bibnamefont {Kumar}},
  \bibinfo {author} {\bibfnamefont {M.~K.}\ \bibnamefont {Verma}}, \ and\
  \bibinfo {author} {\bibfnamefont {S.}~\bibnamefont {Chakraborty}},\
  }\bibfield  {title} {\enquote {\bibinfo {title} {Statistical features of
  rapidly rotating decaying turbulence: Enstrophy and energy spectra and
  coherent structures},}\ }\href@noop {} {\bibfield  {journal} {\bibinfo
  {journal} {Physics of Fluids}\ }\textbf {\bibinfo {volume} {30}},\ \bibinfo
  {pages} {045103} (\bibinfo {year} {2018})}\BibitemShut {NoStop}%
\bibitem [{\citenamefont {Sharma}, \citenamefont {Verma},\ and\ \citenamefont
  {Chakraborty}(2018)}]{Sharma2018}%
  \BibitemOpen
  \bibfield  {author} {\bibinfo {author} {\bibfnamefont {M.~K.}\ \bibnamefont
  {Sharma}}, \bibinfo {author} {\bibfnamefont {M.~K.}\ \bibnamefont {Verma}}, \
  and\ \bibinfo {author} {\bibfnamefont {S.}~\bibnamefont {Chakraborty}},\
  }\bibfield  {title} {\enquote {\bibinfo {title} {On the energy spectrum of
  rapidly rotating forced turbulence},}\ }\href@noop {} {\bibfield  {journal}
  {\bibinfo  {journal} {Physics of Fluids}\ }\textbf {\bibinfo {volume} {30}},\
  \bibinfo {pages} {115102} (\bibinfo {year} {2018})}\BibitemShut {NoStop}%
\bibitem [{\citenamefont {Sharma}, \citenamefont {Verma},\ and\ \citenamefont
  {Chakraborty}(2019)}]{Sharma2019}%
  \BibitemOpen
  \bibfield  {author} {\bibinfo {author} {\bibfnamefont {M.~K.}\ \bibnamefont
  {Sharma}}, \bibinfo {author} {\bibfnamefont {M.~K.}\ \bibnamefont {Verma}}, \
  and\ \bibinfo {author} {\bibfnamefont {S.}~\bibnamefont {Chakraborty}},\
  }\bibfield  {title} {\enquote {\bibinfo {title} {Anisotropic energy transfers
  in rapidly rotating turbulence},}\ }\href@noop {} {\bibfield  {journal}
  {\bibinfo  {journal} {Physics of Fluids}\ }\textbf {\bibinfo {volume} {31}},\
  \bibinfo {pages} {085117} (\bibinfo {year} {2019})}\BibitemShut {NoStop}%
\bibitem [{\citenamefont {Seiwert}, \citenamefont {Morize},\ and\ \citenamefont
  {Moisy}(2008)}]{Seiwert2008}%
  \BibitemOpen
  \bibfield  {author} {\bibinfo {author} {\bibfnamefont {J.}~\bibnamefont
  {Seiwert}}, \bibinfo {author} {\bibfnamefont {C.}~\bibnamefont {Morize}}, \
  and\ \bibinfo {author} {\bibfnamefont {F.}~\bibnamefont {Moisy}},\ }\bibfield
   {title} {\enquote {\bibinfo {title} {On the decrease of intermittency in
  decaying rotating turbulence},}\ }\href@noop {} {\bibfield  {journal}
  {\bibinfo  {journal} {Physics of Fluids}\ }\textbf {\bibinfo {volume} {20}},\
  \bibinfo {pages} {071702} (\bibinfo {year} {2008})}\BibitemShut {NoStop}%
\bibitem [{\citenamefont {Thiele}\ and\ \citenamefont
  {M\"uller}(2009)}]{Thiele2009}%
  \BibitemOpen
  \bibfield  {author} {\bibinfo {author} {\bibfnamefont {M.}~\bibnamefont
  {Thiele}}\ and\ \bibinfo {author} {\bibfnamefont {W.-C.}\ \bibnamefont
  {M\"uller}},\ }\bibfield  {title} {\enquote {\bibinfo {title} {Structure and
  decay of rotating homogeneous turbulence},}\ }\href@noop {} {\bibfield
  {journal} {\bibinfo  {journal} {Journal of Fluid Mechanics}\ }\textbf
  {\bibinfo {volume} {637}},\ \bibinfo {pages} {425} (\bibinfo {year}
  {2009})}\BibitemShut {NoStop}%
\bibitem [{\citenamefont {Mininni}\ and\ \citenamefont
  {Pouquet}(2010{\natexlab{a}})}]{Mininni2010a}%
  \BibitemOpen
  \bibfield  {author} {\bibinfo {author} {\bibfnamefont {P.~D.}\ \bibnamefont
  {Mininni}}\ and\ \bibinfo {author} {\bibfnamefont {A.}~\bibnamefont
  {Pouquet}},\ }\bibfield  {title} {\enquote {\bibinfo {title} {Rotating
  helical turbulence. {II}. intermittency, scale invariance, and structures},}\
  }\href@noop {} {\bibfield  {journal} {\bibinfo  {journal} {Physics of
  Fluids}\ }\textbf {\bibinfo {volume} {22}},\ \bibinfo {pages} {035106}
  (\bibinfo {year} {2010}{\natexlab{a}})}\BibitemShut {NoStop}%
\bibitem [{\citenamefont {Imazio}\ and\ \citenamefont
  {Mininni}(2017)}]{Imazio2017}%
  \BibitemOpen
  \bibfield  {author} {\bibinfo {author} {\bibfnamefont {P.~R.}\ \bibnamefont
  {Imazio}}\ and\ \bibinfo {author} {\bibfnamefont {P.~D.}\ \bibnamefont
  {Mininni}},\ }\bibfield  {title} {\enquote {\bibinfo {title} {Passive
  scalars: Mixing, diffusion, and intermittency in helical and nonhelical
  rotating turbulence},}\ }\href@noop {} {\bibfield  {journal} {\bibinfo
  {journal} {Physical Review E}\ }\textbf {\bibinfo {volume} {95}} (\bibinfo
  {year} {2017})}\BibitemShut {NoStop}%
\bibitem [{\citenamefont {Kraichnan}(1958)}]{Kraichnan1958}%
  \BibitemOpen
  \bibfield  {author} {\bibinfo {author} {\bibfnamefont {R.~H.}\ \bibnamefont
  {Kraichnan}},\ }\bibfield  {title} {\enquote {\bibinfo {title} {Higher order
  interactions in homogeneous turbulence theory},}\ }\href@noop {} {\bibfield
  {journal} {\bibinfo  {journal} {The Physics of Fluids}\ }\textbf {\bibinfo
  {volume} {1}},\ \bibinfo {pages} {358--359} (\bibinfo {year}
  {1958})}\BibitemShut {NoStop}%
\bibitem [{\citenamefont {Kraichnan}(1959)}]{Kraichnan1959}%
  \BibitemOpen
  \bibfield  {author} {\bibinfo {author} {\bibfnamefont {R.~H.}\ \bibnamefont
  {Kraichnan}},\ }\bibfield  {title} {\enquote {\bibinfo {title} {The structure
  of isotropic turbulence at very high reynolds numbers},}\ }\href@noop {}
  {\bibfield  {journal} {\bibinfo  {journal} {Journal of Fluid Mechanics}\
  }\textbf {\bibinfo {volume} {5}},\ \bibinfo {pages} {497–543} (\bibinfo
  {year} {1959})}\BibitemShut {NoStop}%
\bibitem [{\citenamefont {Orszag}(1970)}]{Orszag}%
  \BibitemOpen
  \bibfield  {author} {\bibinfo {author} {\bibfnamefont {S.~A.}\ \bibnamefont
  {Orszag}},\ }\bibfield  {title} {\enquote {\bibinfo {title} {Analytical
  theories of turbulence},}\ }\href@noop {} {\bibfield  {journal} {\bibinfo
  {journal} {Journal of Fluid Mechanics}\ }\textbf {\bibinfo {volume} {41}},\
  \bibinfo {pages} {363–386} (\bibinfo {year} {1970})}\BibitemShut {NoStop}%
\bibitem [{\citenamefont {Lesieur}(2008)}]{Lesieur}%
  \BibitemOpen
  \bibfield  {author} {\bibinfo {author} {\bibfnamefont {M.}~\bibnamefont
  {Lesieur}},\ }\href@noop {} {\emph {\bibinfo {title} {Turbulence in
  Fluids}}}\ (\bibinfo  {publisher} {Springer},\ \bibinfo {year}
  {2008})\BibitemShut {NoStop}%
\bibitem [{\citenamefont {Frisch}\ and\ \citenamefont
  {Bec}(2001)}]{Frisch2001}%
  \BibitemOpen
  \bibfield  {author} {\bibinfo {author} {\bibfnamefont {U.}~\bibnamefont
  {Frisch}}\ and\ \bibinfo {author} {\bibfnamefont {J.}~\bibnamefont {Bec}},\
  }\enquote {\bibinfo {title} {Burgulence},}\ in\ \href@noop {} {\emph
  {\bibinfo {booktitle} {New trends in turbulence Turbulence: nouveaux aspects:
  31 July -- 1 September 2000}}}\ (\bibinfo  {publisher} {Springer Berlin
  Heidelberg},\ \bibinfo {address} {Berlin, Heidelberg},\ \bibinfo {year}
  {2001})\ pp.\ \bibinfo {pages} {341--383}\BibitemShut {NoStop}%
\bibitem [{\citenamefont {Saha}, \citenamefont {Chakraborty},\ and\
  \citenamefont {Bhattacharjee}(2008)}]{Saha2008}%
  \BibitemOpen
  \bibfield  {author} {\bibinfo {author} {\bibfnamefont {A.}~\bibnamefont
  {Saha}}, \bibinfo {author} {\bibfnamefont {S.}~\bibnamefont {Chakraborty}}, \
  and\ \bibinfo {author} {\bibfnamefont {J.}~\bibnamefont {Bhattacharjee}},\
  }\bibfield  {title} {\enquote {\bibinfo {title} {Dominance of rare events in
  some problems in statistical physics},}\ }\href@noop {} {\bibfield  {journal}
  {\bibinfo  {journal} {Pramana}\ }\textbf {\bibinfo {volume} {71}},\ \bibinfo
  {pages} {413--422} (\bibinfo {year} {2008})}\BibitemShut {NoStop}%
\bibitem [{\citenamefont {Chakraborty}, \citenamefont {Saha},\ and\
  \citenamefont {Bhattacharjee}(2009)}]{Chakraborty2009}%
  \BibitemOpen
  \bibfield  {author} {\bibinfo {author} {\bibfnamefont {S.}~\bibnamefont
  {Chakraborty}}, \bibinfo {author} {\bibfnamefont {A.}~\bibnamefont {Saha}}, \
  and\ \bibinfo {author} {\bibfnamefont {J.~K.}\ \bibnamefont
  {Bhattacharjee}},\ }\bibfield  {title} {\enquote {\bibinfo {title} {Large
  deviation theory for coin tossing and turbulence},}\ }\href@noop {}
  {\bibfield  {journal} {\bibinfo  {journal} {Physical Review E}\ }\textbf
  {\bibinfo {volume} {80}} (\bibinfo {year} {2009})}\BibitemShut {NoStop}%
\bibitem [{\citenamefont {Ray}(2015)}]{Ray-Review}%
  \BibitemOpen
  \bibfield  {author} {\bibinfo {author} {\bibfnamefont {S.~S.}\ \bibnamefont
  {Ray}},\ }\bibfield  {title} {\enquote {\bibinfo {title} {Turbulence in
  noninteger dimensions by fractal fourier decimation},}\ }\href@noop {}
  {\bibfield  {journal} {\bibinfo  {journal} {Pramana}\ }\textbf {\bibinfo
  {volume} {84}},\ \bibinfo {pages} {395} (\bibinfo {year} {2015})}\BibitemShut
  {NoStop}%
\bibitem [{\citenamefont {Ray}(2018)}]{Ray-PRF2018}%
  \BibitemOpen
  \bibfield  {author} {\bibinfo {author} {\bibfnamefont {S.~S.}\ \bibnamefont
  {Ray}},\ }\bibfield  {title} {\enquote {\bibinfo {title} {Non-intermittent
  turbulence: Lagrangian chaos and irreversibility},}\ }\href@noop {}
  {\bibfield  {journal} {\bibinfo  {journal} {Phys. Rev. Fluids}\ }\textbf
  {\bibinfo {volume} {3}},\ \bibinfo {pages} {072601} (\bibinfo {year}
  {2018})}\BibitemShut {NoStop}%
\bibitem [{\citenamefont {Lepreti}, \citenamefont {Carbone},\ and\
  \citenamefont {Veltri}(2006)}]{Lepreti2006}%
  \BibitemOpen
  \bibfield  {author} {\bibinfo {author} {\bibfnamefont {F.}~\bibnamefont
  {Lepreti}}, \bibinfo {author} {\bibfnamefont {V.}~\bibnamefont {Carbone}}, \
  and\ \bibinfo {author} {\bibfnamefont {P.}~\bibnamefont {Veltri}},\
  }\bibfield  {title} {\enquote {\bibinfo {title} {Model for intermittency of
  energy dissipation in turbulent flows},}\ }\href@noop {} {\bibfield
  {journal} {\bibinfo  {journal} {Physical Review E}\ }\textbf {\bibinfo
  {volume} {74}} (\bibinfo {year} {2006})}\BibitemShut {NoStop}%
\bibitem [{\citenamefont {Bohr}\ \emph {et~al.}(1998)\citenamefont {Bohr},
  \citenamefont {Jensen}, \citenamefont {Paladin},\ and\ \citenamefont
  {Vulpiani}}]{Bohrbook}%
  \BibitemOpen
  \bibfield  {author} {\bibinfo {author} {\bibfnamefont {T.}~\bibnamefont
  {Bohr}}, \bibinfo {author} {\bibfnamefont {M.~H.}\ \bibnamefont {Jensen}},
  \bibinfo {author} {\bibfnamefont {G.}~\bibnamefont {Paladin}}, \ and\
  \bibinfo {author} {\bibfnamefont {A.}~\bibnamefont {Vulpiani}},\ }\href@noop
  {} {\emph {\bibinfo {title} {Dynamical Systems Approach to Turbulence}}}\
  (\bibinfo  {publisher} {Cambridge University Press},\ \bibinfo {year}
  {1998})\BibitemShut {NoStop}%
\bibitem [{\citenamefont {Biferale}(2003)}]{Biferale-review}%
  \BibitemOpen
  \bibfield  {author} {\bibinfo {author} {\bibfnamefont {L.}~\bibnamefont
  {Biferale}},\ }\bibfield  {title} {\enquote {\bibinfo {title} {Shell models
  of energy cascade in turbulence},}\ }\href@noop {} {\bibfield  {journal}
  {\bibinfo  {journal} {Annual Review of Fluid Mechanics}\ }\textbf {\bibinfo
  {volume} {35}},\ \bibinfo {pages} {441--468} (\bibinfo {year}
  {2003})}\BibitemShut {NoStop}%
\bibitem [{\citenamefont {Biskamp}(1994)}]{Biskamp1994}%
  \BibitemOpen
  \bibfield  {author} {\bibinfo {author} {\bibfnamefont {D.}~\bibnamefont
  {Biskamp}},\ }\bibfield  {title} {\enquote {\bibinfo {title} {Cascade models
  for magnetohydrodynamic turbulence},}\ }\href@noop {} {\bibfield  {journal}
  {\bibinfo  {journal} {Phys. Rev. E}\ }\textbf {\bibinfo {volume} {50}},\
  \bibinfo {pages} {2702--2711} (\bibinfo {year} {1994})}\BibitemShut {NoStop}%
\bibitem [{\citenamefont {Wirth}\ and\ \citenamefont
  {Biferale}(1996)}]{Wirth1996}%
  \BibitemOpen
  \bibfield  {author} {\bibinfo {author} {\bibfnamefont {A.}~\bibnamefont
  {Wirth}}\ and\ \bibinfo {author} {\bibfnamefont {L.}~\bibnamefont
  {Biferale}},\ }\bibfield  {title} {\enquote {\bibinfo {title} {Anomalous
  scaling in random shell models for passive scalars},}\ }\href@noop {}
  {\bibfield  {journal} {\bibinfo  {journal} {Phys. Rev. E}\ }\textbf {\bibinfo
  {volume} {54}},\ \bibinfo {pages} {4982--4989} (\bibinfo {year}
  {1996})}\BibitemShut {NoStop}%
\bibitem [{\citenamefont {Basu}\ \emph {et~al.}(1998)\citenamefont {Basu},
  \citenamefont {Sain}, \citenamefont {Dhar},\ and\ \citenamefont
  {Pandit}}]{Basu1998}%
  \BibitemOpen
  \bibfield  {author} {\bibinfo {author} {\bibfnamefont {A.}~\bibnamefont
  {Basu}}, \bibinfo {author} {\bibfnamefont {A.}~\bibnamefont {Sain}}, \bibinfo
  {author} {\bibfnamefont {S.~K.}\ \bibnamefont {Dhar}}, \ and\ \bibinfo
  {author} {\bibfnamefont {R.}~\bibnamefont {Pandit}},\ }\bibfield  {title}
  {\enquote {\bibinfo {title} {Multiscaling in models of magnetohydrodynamic
  turbulence},}\ }\href@noop {} {\bibfield  {journal} {\bibinfo  {journal}
  {Phys. Rev. Lett.}\ }\textbf {\bibinfo {volume} {81}},\ \bibinfo {pages}
  {2687--2690} (\bibinfo {year} {1998})}\BibitemShut {NoStop}%
\bibitem [{\citenamefont {Frick}\ and\ \citenamefont
  {Sokoloff}(1998)}]{Frick1998}%
  \BibitemOpen
  \bibfield  {author} {\bibinfo {author} {\bibfnamefont {P.}~\bibnamefont
  {Frick}}\ and\ \bibinfo {author} {\bibfnamefont {D.}~\bibnamefont
  {Sokoloff}},\ }\bibfield  {title} {\enquote {\bibinfo {title} {Cascade and
  dynamo action in a shell model of magnetohydrodynamic turbulence},}\
  }\href@noop {} {\bibfield  {journal} {\bibinfo  {journal} {Phys. Rev. E}\
  }\textbf {\bibinfo {volume} {57}},\ \bibinfo {pages} {4155--4164} (\bibinfo
  {year} {1998})}\BibitemShut {NoStop}%
\bibitem [{\citenamefont {Mitra}\ and\ \citenamefont
  {Pandit}(2004)}]{Mitra2004}%
  \BibitemOpen
  \bibfield  {author} {\bibinfo {author} {\bibfnamefont {D.}~\bibnamefont
  {Mitra}}\ and\ \bibinfo {author} {\bibfnamefont {R.}~\bibnamefont {Pandit}},\
  }\bibfield  {title} {\enquote {\bibinfo {title} {Varieties of dynamic
  multiscaling in fluid turbulence},}\ }\href@noop {} {\bibfield  {journal}
  {\bibinfo  {journal} {Phys. Rev. Lett.}\ }\textbf {\bibinfo {volume} {93}},\
  \bibinfo {pages} {024501} (\bibinfo {year} {2004})}\BibitemShut {NoStop}%
\bibitem [{\citenamefont {Mitra}\ and\ \citenamefont
  {Pandit}(2005)}]{Mitra2005}%
  \BibitemOpen
  \bibfield  {author} {\bibinfo {author} {\bibfnamefont {D.}~\bibnamefont
  {Mitra}}\ and\ \bibinfo {author} {\bibfnamefont {R.}~\bibnamefont {Pandit}},\
  }\bibfield  {title} {\enquote {\bibinfo {title} {Dynamics of passive-scalar
  turbulence},}\ }\href@noop {} {\bibfield  {journal} {\bibinfo  {journal}
  {Phys. Rev. Lett.}\ }\textbf {\bibinfo {volume} {95}},\ \bibinfo {pages}
  {144501} (\bibinfo {year} {2005})}\BibitemShut {NoStop}%
\bibitem [{\citenamefont {Ray}, \citenamefont {Mitra},\ and\ \citenamefont
  {Pandit}(2008)}]{Ray2008}%
  \BibitemOpen
  \bibfield  {author} {\bibinfo {author} {\bibfnamefont {S.~S.}\ \bibnamefont
  {Ray}}, \bibinfo {author} {\bibfnamefont {D.}~\bibnamefont {Mitra}}, \ and\
  \bibinfo {author} {\bibfnamefont {R.}~\bibnamefont {Pandit}},\ }\bibfield
  {title} {\enquote {\bibinfo {title} {The universality of dynamic multiscaling
  in homogeneous, isotropic navier{\textendash}stokes and passive-scalar
  turbulence},}\ }\href@noop {} {\bibfield  {journal} {\bibinfo  {journal} {New
  Journal of Physics}\ }\textbf {\bibinfo {volume} {10}},\ \bibinfo {pages}
  {033003} (\bibinfo {year} {2008})}\BibitemShut {NoStop}%
\bibitem [{\citenamefont {Ray}\ and\ \citenamefont {Basu}(2011)}]{Ray2011}%
  \BibitemOpen
  \bibfield  {author} {\bibinfo {author} {\bibfnamefont {S.~S.}\ \bibnamefont
  {Ray}}\ and\ \bibinfo {author} {\bibfnamefont {A.}~\bibnamefont {Basu}},\
  }\bibfield  {title} {\enquote {\bibinfo {title} {Universality of scaling and
  multiscaling in turbulent symmetric binary fluids},}\ }\href@noop {}
  {\bibfield  {journal} {\bibinfo  {journal} {Phys. Rev. E}\ }\textbf {\bibinfo
  {volume} {84}},\ \bibinfo {pages} {036316} (\bibinfo {year}
  {2011})}\BibitemShut {NoStop}%
\bibitem [{\citenamefont {Banerjee}\ \emph {et~al.}(2013)\citenamefont
  {Banerjee}, \citenamefont {Ray}, \citenamefont {Sahoo},\ and\ \citenamefont
  {Pandit}}]{Banerjee2013}%
  \BibitemOpen
  \bibfield  {author} {\bibinfo {author} {\bibfnamefont {D.}~\bibnamefont
  {Banerjee}}, \bibinfo {author} {\bibfnamefont {S.~S.}\ \bibnamefont {Ray}},
  \bibinfo {author} {\bibfnamefont {G.}~\bibnamefont {Sahoo}}, \ and\ \bibinfo
  {author} {\bibfnamefont {R.}~\bibnamefont {Pandit}},\ }\bibfield  {title}
  {\enquote {\bibinfo {title} {Multiscaling in hall-magnetohydrodynamic
  turbulence: Insights from a shell model},}\ }\href@noop {} {\bibfield
  {journal} {\bibinfo  {journal} {Phys. Rev. Lett.}\ }\textbf {\bibinfo
  {volume} {111}},\ \bibinfo {pages} {174501} (\bibinfo {year}
  {2013})}\BibitemShut {NoStop}%
\bibitem [{\citenamefont {Benzi}\ \emph {et~al.}(2003)\citenamefont {Benzi},
  \citenamefont {De~Angelis}, \citenamefont {Govindarajan},\ and\ \citenamefont
  {Procaccia}}]{Benzi2003}%
  \BibitemOpen
  \bibfield  {author} {\bibinfo {author} {\bibfnamefont {R.}~\bibnamefont
  {Benzi}}, \bibinfo {author} {\bibfnamefont {E.}~\bibnamefont {De~Angelis}},
  \bibinfo {author} {\bibfnamefont {R.}~\bibnamefont {Govindarajan}}, \ and\
  \bibinfo {author} {\bibfnamefont {I.}~\bibnamefont {Procaccia}},\ }\bibfield
  {title} {\enquote {\bibinfo {title} {Shell model for drag reduction with
  polymer additives in homogeneous turbulence},}\ }\href@noop {} {\bibfield
  {journal} {\bibinfo  {journal} {Phys. Rev. E}\ }\textbf {\bibinfo {volume}
  {68}},\ \bibinfo {pages} {016308} (\bibinfo {year} {2003})}\BibitemShut
  {NoStop}%
\bibitem [{\citenamefont {Kalelkar}, \citenamefont {Govindarajan},\ and\
  \citenamefont {Pandit}(2005)}]{Kalelkar2005}%
  \BibitemOpen
  \bibfield  {author} {\bibinfo {author} {\bibfnamefont {C.}~\bibnamefont
  {Kalelkar}}, \bibinfo {author} {\bibfnamefont {R.}~\bibnamefont
  {Govindarajan}}, \ and\ \bibinfo {author} {\bibfnamefont {R.}~\bibnamefont
  {Pandit}},\ }\bibfield  {title} {\enquote {\bibinfo {title} {Drag reduction
  by polymer additives in decaying turbulence},}\ }\href@noop {} {\bibfield
  {journal} {\bibinfo  {journal} {Phys. Rev. E}\ }\textbf {\bibinfo {volume}
  {72}},\ \bibinfo {pages} {017301} (\bibinfo {year} {2005})}\BibitemShut
  {NoStop}%
\bibitem [{\citenamefont {Ray}\ and\ \citenamefont {Vincenzi}(2016)}]{Ray2016}%
  \BibitemOpen
  \bibfield  {author} {\bibinfo {author} {\bibfnamefont {S.~S.}\ \bibnamefont
  {Ray}}\ and\ \bibinfo {author} {\bibfnamefont {D.}~\bibnamefont {Vincenzi}},\
  }\bibfield  {title} {\enquote {\bibinfo {title} {Elastic turbulence in a
  shell model of polymer solution},}\ }\href@noop {} {\bibfield  {journal}
  {\bibinfo  {journal} {{EPL} (Europhysics Letters)}\ }\textbf {\bibinfo
  {volume} {114}},\ \bibinfo {pages} {44001} (\bibinfo {year}
  {2016})}\BibitemShut {NoStop}%
\bibitem [{\citenamefont {Ditlevsen}\ and\ \citenamefont
  {Mogensen}(1996)}]{Ditlevsen1996}%
  \BibitemOpen
  \bibfield  {author} {\bibinfo {author} {\bibfnamefont {P.~D.}\ \bibnamefont
  {Ditlevsen}}\ and\ \bibinfo {author} {\bibfnamefont {I.~A.}\ \bibnamefont
  {Mogensen}},\ }\bibfield  {title} {\enquote {\bibinfo {title} {Cascades and
  statistical equilibrium in shell models of turbulence},}\ }\href@noop {}
  {\bibfield  {journal} {\bibinfo  {journal} {Phys. Rev. E}\ }\textbf {\bibinfo
  {volume} {53}},\ \bibinfo {pages} {4785--4793} (\bibinfo {year}
  {1996})}\BibitemShut {NoStop}%
\bibitem [{\citenamefont {Gilbert}\ \emph {et~al.}(2002)\citenamefont
  {Gilbert}, \citenamefont {L'vov}, \citenamefont {Pomyalov},\ and\
  \citenamefont {Procaccia}}]{Gilbert2002}%
  \BibitemOpen
  \bibfield  {author} {\bibinfo {author} {\bibfnamefont {T.}~\bibnamefont
  {Gilbert}}, \bibinfo {author} {\bibfnamefont {V.~S.}\ \bibnamefont {L'vov}},
  \bibinfo {author} {\bibfnamefont {A.}~\bibnamefont {Pomyalov}}, \ and\
  \bibinfo {author} {\bibfnamefont {I.}~\bibnamefont {Procaccia}},\ }\bibfield
  {title} {\enquote {\bibinfo {title} {Inverse cascade regime in shell models
  of two-dimensional turbulence},}\ }\href@noop {} {\bibfield  {journal}
  {\bibinfo  {journal} {Phys. Rev. Lett.}\ }\textbf {\bibinfo {volume} {89}},\
  \bibinfo {pages} {074501} (\bibinfo {year} {2002})}\BibitemShut {NoStop}%
\bibitem [{\citenamefont {Tom}\ and\ \citenamefont {Ray}(2017)}]{Tom2017}%
  \BibitemOpen
  \bibfield  {author} {\bibinfo {author} {\bibfnamefont {R.}~\bibnamefont
  {Tom}}\ and\ \bibinfo {author} {\bibfnamefont {S.~S.}\ \bibnamefont {Ray}},\
  }\bibfield  {title} {\enquote {\bibinfo {title} {Revisiting the {SABRA}
  model: Statics and dynamics},}\ }\href@noop {} {\bibfield  {journal}
  {\bibinfo  {journal} {{EPL} (Europhysics Letters)}\ }\textbf {\bibinfo
  {volume} {120}},\ \bibinfo {pages} {34002} (\bibinfo {year}
  {2017})}\BibitemShut {NoStop}%
\bibitem [{\citenamefont {Chakraborty}, \citenamefont {Jensen},\ and\
  \citenamefont {Sarkar}(2010)}]{Chakraborty2010}%
  \BibitemOpen
  \bibfield  {author} {\bibinfo {author} {\bibfnamefont {S.}~\bibnamefont
  {Chakraborty}}, \bibinfo {author} {\bibfnamefont {M.~H.}\ \bibnamefont
  {Jensen}}, \ and\ \bibinfo {author} {\bibfnamefont {A.}~\bibnamefont
  {Sarkar}},\ }\bibfield  {title} {\enquote {\bibinfo {title} {On
  two-dimensionalization of three-dimensional turbulence in shell models},}\
  }\href@noop {} {\bibfield  {journal} {\bibinfo  {journal} {The European
  Physical Journal B}\ }\textbf {\bibinfo {volume} {73}},\ \bibinfo {pages}
  {447--453} (\bibinfo {year} {2010})}\BibitemShut {NoStop}%
\bibitem [{\citenamefont {Benzi}\ \emph {et~al.}(1996)\citenamefont {Benzi},
  \citenamefont {Biferale}, \citenamefont {Kerr},\ and\ \citenamefont
  {Trovatore}}]{Benzi1996}%
  \BibitemOpen
  \bibfield  {author} {\bibinfo {author} {\bibfnamefont {R.}~\bibnamefont
  {Benzi}}, \bibinfo {author} {\bibfnamefont {L.}~\bibnamefont {Biferale}},
  \bibinfo {author} {\bibfnamefont {R.~M.}\ \bibnamefont {Kerr}}, \ and\
  \bibinfo {author} {\bibfnamefont {E.}~\bibnamefont {Trovatore}},\ }\bibfield
  {title} {\enquote {\bibinfo {title} {Helical shell models for
  three-dimensional turbulence},}\ }\href@noop {} {\bibfield  {journal}
  {\bibinfo  {journal} {Physical Review E}\ }\textbf {\bibinfo {volume} {53}},\
  \bibinfo {pages} {3541--3550} (\bibinfo {year} {1996})}\BibitemShut {NoStop}%
\bibitem [{\citenamefont {Waleffe}(1992)}]{Waleffe1992}%
  \BibitemOpen
  \bibfield  {author} {\bibinfo {author} {\bibfnamefont {F.}~\bibnamefont
  {Waleffe}},\ }\bibfield  {title} {\enquote {\bibinfo {title} {The nature of
  triad interactions in homogeneous turbulence},}\ }\href@noop {} {\bibfield
  {journal} {\bibinfo  {journal} {Physics of Fluids A: Fluid Dynamics}\
  }\textbf {\bibinfo {volume} {4}},\ \bibinfo {pages} {350--363} (\bibinfo
  {year} {1992})}\BibitemShut {NoStop}%
\bibitem [{\citenamefont {Zhou}(1995)}]{Zhou1995}%
  \BibitemOpen
  \bibfield  {author} {\bibinfo {author} {\bibfnamefont {Y.}~\bibnamefont
  {Zhou}},\ }\bibfield  {title} {\enquote {\bibinfo {title} {A phenomenological
  treatment of rotating turbulence},}\ }\href@noop {} {\bibfield  {journal}
  {\bibinfo  {journal} {Physics of Fluids}\ }\textbf {\bibinfo {volume} {7}},\
  \bibinfo {pages} {2092--2094} (\bibinfo {year} {1995})}\BibitemShut {NoStop}%
\bibitem [{\citenamefont {Yeung}\ and\ \citenamefont {Zhou}(1998)}]{Yeung1998}%
  \BibitemOpen
  \bibfield  {author} {\bibinfo {author} {\bibfnamefont {P.~K.}\ \bibnamefont
  {Yeung}}\ and\ \bibinfo {author} {\bibfnamefont {Y.}~\bibnamefont {Zhou}},\
  }\bibfield  {title} {\enquote {\bibinfo {title} {Numerical study of rotating
  turbulence with external forcing},}\ }\href@noop {} {\bibfield  {journal}
  {\bibinfo  {journal} {Physics of Fluids}\ }\textbf {\bibinfo {volume} {10}},\
  \bibinfo {pages} {2895--2909} (\bibinfo {year} {1998})}\BibitemShut {NoStop}%
\bibitem [{\citenamefont {Baroud}\ \emph {et~al.}(2002)\citenamefont {Baroud},
  \citenamefont {Plapp}, \citenamefont {She},\ and\ \citenamefont
  {Swinney}}]{Baroud2002}%
  \BibitemOpen
  \bibfield  {author} {\bibinfo {author} {\bibfnamefont {C.~N.}\ \bibnamefont
  {Baroud}}, \bibinfo {author} {\bibfnamefont {B.~B.}\ \bibnamefont {Plapp}},
  \bibinfo {author} {\bibfnamefont {Z.-S.}\ \bibnamefont {She}}, \ and\
  \bibinfo {author} {\bibfnamefont {H.~L.}\ \bibnamefont {Swinney}},\
  }\bibfield  {title} {\enquote {\bibinfo {title} {Anomalous self-similarity in
  a turbulent rapidly rotating fluid},}\ }\href@noop {} {\bibfield  {journal}
  {\bibinfo  {journal} {Physical Review Letters}\ }\textbf {\bibinfo {volume}
  {88}} (\bibinfo {year} {2002})}\BibitemShut {NoStop}%
\bibitem [{\citenamefont {Baroud}\ \emph {et~al.}(2003)\citenamefont {Baroud},
  \citenamefont {Plapp}, \citenamefont {Swinney},\ and\ \citenamefont
  {She}}]{Baroud2003}%
  \BibitemOpen
  \bibfield  {author} {\bibinfo {author} {\bibfnamefont {C.~N.}\ \bibnamefont
  {Baroud}}, \bibinfo {author} {\bibfnamefont {B.~B.}\ \bibnamefont {Plapp}},
  \bibinfo {author} {\bibfnamefont {H.~L.}\ \bibnamefont {Swinney}}, \ and\
  \bibinfo {author} {\bibfnamefont {Z.-S.}\ \bibnamefont {She}},\ }\bibfield
  {title} {\enquote {\bibinfo {title} {Scaling in three-dimensional and
  quasi-two-dimensional rotating turbulent flows},}\ }\href@noop {} {\bibfield
  {journal} {\bibinfo  {journal} {Physics of Fluids}\ }\textbf {\bibinfo
  {volume} {15}},\ \bibinfo {pages} {2091--2104} (\bibinfo {year}
  {2003})}\BibitemShut {NoStop}%
\bibitem [{\citenamefont {Mininni}\ and\ \citenamefont
  {Pouquet}(2010{\natexlab{b}})}]{Mininni2010}%
  \BibitemOpen
  \bibfield  {author} {\bibinfo {author} {\bibfnamefont {P.~D.}\ \bibnamefont
  {Mininni}}\ and\ \bibinfo {author} {\bibfnamefont {A.}~\bibnamefont
  {Pouquet}},\ }\bibfield  {title} {\enquote {\bibinfo {title} {Rotating
  helical turbulence. {I}. global evolution and spectral behavior},}\
  }\href@noop {} {\bibfield  {journal} {\bibinfo  {journal} {Physics of
  Fluids}\ }\textbf {\bibinfo {volume} {22}},\ \bibinfo {pages} {035105}
  (\bibinfo {year} {2010}{\natexlab{b}})}\BibitemShut {NoStop}%
\bibitem [{\citenamefont {Benzi}\ \emph {et~al.}(1993)\citenamefont {Benzi},
  \citenamefont {Ciliberto}, \citenamefont {Tripiccione}, \citenamefont
  {Baudet}, \citenamefont {Massaioli},\ and\ \citenamefont
  {Succi}}]{Benzi1993}%
  \BibitemOpen
  \bibfield  {author} {\bibinfo {author} {\bibfnamefont {R.}~\bibnamefont
  {Benzi}}, \bibinfo {author} {\bibfnamefont {S.}~\bibnamefont {Ciliberto}},
  \bibinfo {author} {\bibfnamefont {R.}~\bibnamefont {Tripiccione}}, \bibinfo
  {author} {\bibfnamefont {C.}~\bibnamefont {Baudet}}, \bibinfo {author}
  {\bibfnamefont {F.}~\bibnamefont {Massaioli}}, \ and\ \bibinfo {author}
  {\bibfnamefont {S.}~\bibnamefont {Succi}},\ }\bibfield  {title} {\enquote
  {\bibinfo {title} {Extended self-similarity in turbulent flows},}\
  }\href@noop {} {\bibfield  {journal} {\bibinfo  {journal} {Physical Review
  E}\ }\textbf {\bibinfo {volume} {48}},\ \bibinfo {pages} {R29--R32} (\bibinfo
  {year} {1993})}\BibitemShut {NoStop}%
\bibitem [{\citenamefont {Chakraborty}, \citenamefont {Frisch},\ and\
  \citenamefont {Ray}(2010)}]{Chakraborty2010a}%
  \BibitemOpen
  \bibfield  {author} {\bibinfo {author} {\bibfnamefont {S.}~\bibnamefont
  {Chakraborty}}, \bibinfo {author} {\bibfnamefont {U.}~\bibnamefont {Frisch}},
  \ and\ \bibinfo {author} {\bibfnamefont {S.~S.}\ \bibnamefont {Ray}},\
  }\bibfield  {title} {\enquote {\bibinfo {title} {Extended self-similarity
  works for the burgers equation and why},}\ }\href@noop {} {\bibfield
  {journal} {\bibinfo  {journal} {Journal of Fluid Mechanics}\ }\textbf
  {\bibinfo {volume} {649}},\ \bibinfo {pages} {275--285} (\bibinfo {year}
  {2010})}\BibitemShut {NoStop}%
\bibitem [{\citenamefont {Chakraborty}\ \emph {et~al.}(2012)\citenamefont
  {Chakraborty}, \citenamefont {Frisch}, \citenamefont {Pauls},\ and\
  \citenamefont {Ray}}]{Chakraborty2012}%
  \BibitemOpen
  \bibfield  {author} {\bibinfo {author} {\bibfnamefont {S.}~\bibnamefont
  {Chakraborty}}, \bibinfo {author} {\bibfnamefont {U.}~\bibnamefont {Frisch}},
  \bibinfo {author} {\bibfnamefont {W.}~\bibnamefont {Pauls}}, \ and\ \bibinfo
  {author} {\bibfnamefont {S.~S.}\ \bibnamefont {Ray}},\ }\bibfield  {title}
  {\enquote {\bibinfo {title} {Nelkin scaling for the burgers equation and the
  role of high-precision calculations},}\ }\href@noop {} {\bibfield  {journal}
  {\bibinfo  {journal} {Physical Review E}\ }\textbf {\bibinfo {volume} {85}}
  (\bibinfo {year} {2012})}\BibitemShut {NoStop}%
\bibitem [{\citenamefont {Meneveau}\ and\ \citenamefont
  {Sreenivasan}(1987{\natexlab{a}})}]{Meneveau1987}%
  \BibitemOpen
  \bibfield  {author} {\bibinfo {author} {\bibfnamefont {C.}~\bibnamefont
  {Meneveau}}\ and\ \bibinfo {author} {\bibfnamefont {K.~R.}\ \bibnamefont
  {Sreenivasan}},\ }\bibfield  {title} {\enquote {\bibinfo {title} {Simple
  multifractal cascade model for fully developed turbulence},}\ }\href@noop {}
  {\bibfield  {journal} {\bibinfo  {journal} {Physical Review Letters}\
  }\textbf {\bibinfo {volume} {59}},\ \bibinfo {pages} {1424--1427} (\bibinfo
  {year} {1987}{\natexlab{a}})}\BibitemShut {NoStop}%
\bibitem [{\citenamefont {Hentschel}\ and\ \citenamefont
  {Procaccia}(1983)}]{Hentschel1983}%
  \BibitemOpen
  \bibfield  {author} {\bibinfo {author} {\bibfnamefont {H.}~\bibnamefont
  {Hentschel}}\ and\ \bibinfo {author} {\bibfnamefont {I.}~\bibnamefont
  {Procaccia}},\ }\bibfield  {title} {\enquote {\bibinfo {title} {The infinite
  number of generalized dimensions of fractals and strange attractors},}\
  }\href@noop {} {\bibfield  {journal} {\bibinfo  {journal} {Physica D:
  Nonlinear Phenomena}\ }\textbf {\bibinfo {volume} {8}},\ \bibinfo {pages}
  {435--444} (\bibinfo {year} {1983})}\BibitemShut {NoStop}%
\bibitem [{\citenamefont {Halsey}\ \emph {et~al.}(1986)\citenamefont {Halsey},
  \citenamefont {Jensen}, \citenamefont {Kadanoff}, \citenamefont {Procaccia},\
  and\ \citenamefont {Shraiman}}]{Halsey1986}%
  \BibitemOpen
  \bibfield  {author} {\bibinfo {author} {\bibfnamefont {T.~C.}\ \bibnamefont
  {Halsey}}, \bibinfo {author} {\bibfnamefont {M.~H.}\ \bibnamefont {Jensen}},
  \bibinfo {author} {\bibfnamefont {L.~P.}\ \bibnamefont {Kadanoff}}, \bibinfo
  {author} {\bibfnamefont {I.}~\bibnamefont {Procaccia}}, \ and\ \bibinfo
  {author} {\bibfnamefont {B.~I.}\ \bibnamefont {Shraiman}},\ }\bibfield
  {title} {\enquote {\bibinfo {title} {Fractal measures and their
  singularities: The characterization of strange sets},}\ }\href@noop {}
  {\bibfield  {journal} {\bibinfo  {journal} {Physical Review A}\ }\textbf
  {\bibinfo {volume} {33}},\ \bibinfo {pages} {1141--1151} (\bibinfo {year}
  {1986})}\BibitemShut {NoStop}%
\bibitem [{\citenamefont {Ott}(2002)}]{Ott2002}%
  \BibitemOpen
  \bibfield  {author} {\bibinfo {author} {\bibfnamefont {E.}~\bibnamefont
  {Ott}},\ }\href@noop {} {\emph {\bibinfo {title} {{C}haos in {D}ynamical
  {S}ystems}}}\ (\bibinfo  {publisher} {Cambridge University Press},\ \bibinfo
  {year} {2002})\BibitemShut {NoStop}%
\bibitem [{\citenamefont {Meneveau}\ and\ \citenamefont
  {Sreenivasan}(1991)}]{Meneveau1991}%
  \BibitemOpen
  \bibfield  {author} {\bibinfo {author} {\bibfnamefont {C.}~\bibnamefont
  {Meneveau}}\ and\ \bibinfo {author} {\bibfnamefont {K.~R.}\ \bibnamefont
  {Sreenivasan}},\ }\bibfield  {title} {\enquote {\bibinfo {title} {The
  multifractal nature of turbulent energy dissipation},}\ }\href@noop {}
  {\bibfield  {journal} {\bibinfo  {journal} {Journal of Fluid Mechanics}\
  }\textbf {\bibinfo {volume} {224}},\ \bibinfo {pages} {429} (\bibinfo {year}
  {1991})}\BibitemShut {NoStop}%
\bibitem [{\citenamefont {Chhabra}\ and\ \citenamefont
  {Jensen}(1989)}]{Chhabra1989a}%
  \BibitemOpen
  \bibfield  {author} {\bibinfo {author} {\bibfnamefont {A.}~\bibnamefont
  {Chhabra}}\ and\ \bibinfo {author} {\bibfnamefont {R.~V.}\ \bibnamefont
  {Jensen}},\ }\bibfield  {title} {\enquote {\bibinfo {title} {Direct
  determination of the f($\alpha$) singularity spectrum},}\ }\href@noop {}
  {\bibfield  {journal} {\bibinfo  {journal} {Physical Review Letters}\
  }\textbf {\bibinfo {volume} {62}},\ \bibinfo {pages} {1327--1330} (\bibinfo
  {year} {1989})}\BibitemShut {NoStop}%
\bibitem [{\citenamefont {Chhabra}\ \emph {et~al.}(1989)\citenamefont
  {Chhabra}, \citenamefont {Meneveau}, \citenamefont {Jensen},\ and\
  \citenamefont {Sreenivasan}}]{Chhabra1989}%
  \BibitemOpen
  \bibfield  {author} {\bibinfo {author} {\bibfnamefont {A.~B.}\ \bibnamefont
  {Chhabra}}, \bibinfo {author} {\bibfnamefont {C.}~\bibnamefont {Meneveau}},
  \bibinfo {author} {\bibfnamefont {R.~V.}\ \bibnamefont {Jensen}}, \ and\
  \bibinfo {author} {\bibfnamefont {K.~R.}\ \bibnamefont {Sreenivasan}},\
  }\bibfield  {title} {\enquote {\bibinfo {title} {Direct determination of the
  f($\alpha$) singularity spectrum and its application to fully developed
  turbulence},}\ }\href@noop {} {\bibfield  {journal} {\bibinfo  {journal}
  {Physical Review A}\ }\textbf {\bibinfo {volume} {40}},\ \bibinfo {pages}
  {5284--5294} (\bibinfo {year} {1989})}\BibitemShut {NoStop}%
\bibitem [{\citenamefont {Meneveau}\ and\ \citenamefont
  {Sreenivasan}(1987{\natexlab{b}})}]{Meneveau1987a}%
  \BibitemOpen
  \bibfield  {author} {\bibinfo {author} {\bibfnamefont {C.}~\bibnamefont
  {Meneveau}}\ and\ \bibinfo {author} {\bibfnamefont {K.}~\bibnamefont
  {Sreenivasan}},\ }\bibfield  {title} {\enquote {\bibinfo {title} {The
  multifractal spectrum of the dissipation field in turbulent flows},}\
  }\href@noop {} {\bibfield  {journal} {\bibinfo  {journal} {Nuclear Physics B
  - Proceedings Supplements}\ }\textbf {\bibinfo {volume} {2}},\ \bibinfo
  {pages} {49--76} (\bibinfo {year} {1987}{\natexlab{b}})}\BibitemShut
  {NoStop}%
\bibitem [{\citenamefont {Jensen}(1999)}]{Jensen1999}%
  \BibitemOpen
  \bibfield  {author} {\bibinfo {author} {\bibfnamefont {M.~H.}\ \bibnamefont
  {Jensen}},\ }\bibfield  {title} {\enquote {\bibinfo {title} {Multiscaling and
  structure functions in turbulence: An alternative approach},}\ }\href@noop {}
  {\bibfield  {journal} {\bibinfo  {journal} {Physical Review Letters}\
  }\textbf {\bibinfo {volume} {83}},\ \bibinfo {pages} {76--79} (\bibinfo
  {year} {1999})}\BibitemShut {NoStop}%
\bibitem [{\citenamefont {Roux}\ and\ \citenamefont {Jensen}(2004)}]{Roux2004}%
  \BibitemOpen
  \bibfield  {author} {\bibinfo {author} {\bibfnamefont {S.}~\bibnamefont
  {Roux}}\ and\ \bibinfo {author} {\bibfnamefont {M.~H.}\ \bibnamefont
  {Jensen}},\ }\bibfield  {title} {\enquote {\bibinfo {title} {Dual
  multifractal spectra},}\ }\href@noop {} {\bibfield  {journal} {\bibinfo
  {journal} {Physical Review E}\ }\textbf {\bibinfo {volume} {69}} (\bibinfo
  {year} {2004})}\BibitemShut {NoStop}%
\bibitem [{\citenamefont {Chakraborty}, \citenamefont {Jensen},\ and\
  \citenamefont {Madsen}(2010)}]{Chakraborty2010b}%
  \BibitemOpen
  \bibfield  {author} {\bibinfo {author} {\bibfnamefont {S.}~\bibnamefont
  {Chakraborty}}, \bibinfo {author} {\bibfnamefont {M.~H.}\ \bibnamefont
  {Jensen}}, \ and\ \bibinfo {author} {\bibfnamefont {B.~S.}\ \bibnamefont
  {Madsen}},\ }\bibfield  {title} {\enquote {\bibinfo {title}
  {Three-dimensional turbulent relative dispersion by the
  gledzer-ohkitani-yamada shell model},}\ }\href@noop {} {\bibfield  {journal}
  {\bibinfo  {journal} {Physical Review E}\ }\textbf {\bibinfo {volume} {81}}
  (\bibinfo {year} {2010})}\BibitemShut {NoStop}%
\bibitem [{\citenamefont {Meneveau}(1991)}]{Meneveau1991a}%
  \BibitemOpen
  \bibfield  {author} {\bibinfo {author} {\bibfnamefont {C.}~\bibnamefont
  {Meneveau}},\ }\bibfield  {title} {\enquote {\bibinfo {title} {Analysis of
  turbulence in the orthonormal wavelet representation},}\ }\href@noop {}
  {\bibfield  {journal} {\bibinfo  {journal} {Journal of Fluid Mechanics}\
  }\textbf {\bibinfo {volume} {232}},\ \bibinfo {pages} {469} (\bibinfo {year}
  {1991})}\BibitemShut {NoStop}%
\bibitem [{\citenamefont {Jaffard}, \citenamefont {Lashermes},\ and\
  \citenamefont {Abry}(2007)}]{Jaffard2007}%
  \BibitemOpen
  \bibfield  {author} {\bibinfo {author} {\bibfnamefont {S.}~\bibnamefont
  {Jaffard}}, \bibinfo {author} {\bibfnamefont {B.}~\bibnamefont {Lashermes}},
  \ and\ \bibinfo {author} {\bibfnamefont {P.}~\bibnamefont {Abry}},\
  }\bibfield  {title} {\enquote {\bibinfo {title} {Wavelet leaders in
  multifractal analysis},}\ }in\ \href@noop {} {\emph {\bibinfo {booktitle}
  {Wavelet Analysis and Applications}}},\ \bibinfo {editor} {edited by\
  \bibinfo {editor} {\bibfnamefont {T.}~\bibnamefont {Qian}}, \bibinfo {editor}
  {\bibfnamefont {M.~I.}\ \bibnamefont {Vai}}, \ and\ \bibinfo {editor}
  {\bibfnamefont {Y.}~\bibnamefont {Xu}}}\ (\bibinfo  {publisher} {Birkhäuser
  Basel},\ \bibinfo {year} {2007})\ pp.\ \bibinfo {pages}
  {201--246}\BibitemShut {NoStop}%
\bibitem [{\citenamefont {Wendt}\ and\ \citenamefont
  {Abry}(2007)}]{Wendt_2007}%
  \BibitemOpen
  \bibfield  {author} {\bibinfo {author} {\bibfnamefont {H.}~\bibnamefont
  {Wendt}}\ and\ \bibinfo {author} {\bibfnamefont {P.}~\bibnamefont {Abry}},\
  }\bibfield  {title} {\enquote {\bibinfo {title} {Multifractality tests using
  bootstrapped wavelet leaders},}\ }\href@noop {} {\bibfield  {journal}
  {\bibinfo  {journal} {{IEEE} Transactions on Signal Processing}\ }\textbf
  {\bibinfo {volume} {55}},\ \bibinfo {pages} {4811--4820} (\bibinfo {year}
  {2007})}\BibitemShut {NoStop}%
\bibitem [{\citenamefont {Verma}\ \emph {et~al.}(2013)\citenamefont {Verma},
  \citenamefont {Chatterjee}, \citenamefont {Reddy}, \citenamefont {Yadav},
  \citenamefont {Paul}, \citenamefont {Chandra},\ and\ \citenamefont
  {Samtaney}}]{Verma2013}%
  \BibitemOpen
  \bibfield  {author} {\bibinfo {author} {\bibfnamefont {M.~K.}\ \bibnamefont
  {Verma}}, \bibinfo {author} {\bibfnamefont {A.}~\bibnamefont {Chatterjee}},
  \bibinfo {author} {\bibfnamefont {K.~S.}\ \bibnamefont {Reddy}}, \bibinfo
  {author} {\bibfnamefont {R.~K.}\ \bibnamefont {Yadav}}, \bibinfo {author}
  {\bibfnamefont {S.}~\bibnamefont {Paul}}, \bibinfo {author} {\bibfnamefont
  {M.}~\bibnamefont {Chandra}}, \ and\ \bibinfo {author} {\bibfnamefont
  {R.}~\bibnamefont {Samtaney}},\ }\bibfield  {title} {\enquote {\bibinfo
  {title} {Benchmarking and scaling studies of pseudospectral code tarang for
  turbulence simulations},}\ }\href@noop {} {\bibfield  {journal} {\bibinfo
  {journal} {Pramana}\ }\textbf {\bibinfo {volume} {81}},\ \bibinfo {pages}
  {617--629} (\bibinfo {year} {2013})}\BibitemShut {NoStop}%
\bibitem [{\citenamefont {Chatterjee}\ \emph {et~al.}(2018)\citenamefont
  {Chatterjee}, \citenamefont {Verma}, \citenamefont {Kumar}, \citenamefont
  {Samtaney}, \citenamefont {Hadri},\ and\ \citenamefont
  {Khurram}}]{Chatterjee2018}%
  \BibitemOpen
  \bibfield  {author} {\bibinfo {author} {\bibfnamefont {A.~G.}\ \bibnamefont
  {Chatterjee}}, \bibinfo {author} {\bibfnamefont {M.~K.}\ \bibnamefont
  {Verma}}, \bibinfo {author} {\bibfnamefont {A.}~\bibnamefont {Kumar}},
  \bibinfo {author} {\bibfnamefont {R.}~\bibnamefont {Samtaney}}, \bibinfo
  {author} {\bibfnamefont {B.}~\bibnamefont {Hadri}}, \ and\ \bibinfo {author}
  {\bibfnamefont {R.}~\bibnamefont {Khurram}},\ }\bibfield  {title} {\enquote
  {\bibinfo {title} {Scaling of a fast fourier transform and a pseudo-spectral
  fluid solver up to 196608 cores},}\ }\href@noop {} {\bibfield  {journal}
  {\bibinfo  {journal} {Journal of Parallel and Distributed Computing}\
  }\textbf {\bibinfo {volume} {113}},\ \bibinfo {pages} {77--91} (\bibinfo
  {year} {2018})}\BibitemShut {NoStop}%
\end{thebibliography}%
\end{document}